\pdfoutput=1	% for arXiv % don't have this in the first line when submitting!

\documentclass[final,5p,times,twocolumn]{elsarticle}

\usepackage{graphicx}

\newcounter{bla}

\journal{Computer Physics Communications}

\newcommand{\rmi}{\mathrm{i}}
\newcommand{\rmd}{\mathrm{d}}
\newcommand{\picturewidth}{8.75cm}

\begin{document}

\begin{frontmatter}



\title{TIM, a ray-tracing program for forbidden ray optics}

\author{Dean Lambert$^1$}
\author{Alasdair C.\ Hamilton}
\author{George Constable}
\author{Harsh Snehanshu$^2$}
\author{Sharvil Talati$^2$}
\author{Johannes Courtial\corref{author}}

\cortext[author]{Corresponding author\\\textit{E-mail address:} johannes.courtial@glasgow.ac.uk}
\address{SUPA, School of Physics \& Astronomy, University of Glasgow, Glasgow G12~8QQ, United~Kingdom}

%\author{Alasdair C.\ Hamilton, Dean Lambert, George Constable, and Johannes Courtial}
%\address{SUPA, School of Physics and Astronomy, University of Glasgow, Glasgow G12~8QQ, United~Kingdom}
%\ead{johannes.courtial@glasgow.ac.uk}

\begin{abstract}
TIM (The Interactive METATOY) is a ray-tracing program specifically tailored towards our research in METATOYs, which are optical components that appear to be able to create wave-optically forbidden light-ray fields.
For this reason, TIM possesses features not found in other ray-tracing programs.
TIM can either be used interactively or by modifying the openly available source code; in both cases, it can easily be run as an applet embedded in a web page.
Here we describe the basic structure of TIM's source code and how to extend it, and we give examples of how we have used TIM in our own research.
\end{abstract}

\begin{keyword}
ray tracing; geometrical optics; METATOYs
\end{keyword}

\end{frontmatter}

\footnotetext[1]{Now at SUPA, School of Physics \& Astronomy, University of Edinburgh,
Edinburgh EH9~3JZ, United~Kingdom.}
\footnotetext[2]{HS and ST contributed while visiting from the Indian Institute of Technology Delhi, Hauz Khas, New Delhi 110~016, India.}

%\submitto{}
%
%
%\pacs{
%01.50.Wg, % (Physics of toys)
%42.15.-i, % (Geometrical optics)
%42.15.Dp, % (Wave fronts and ray tracing)
%42.25.Gy, % (Edge and boundary effects; reflection and refraction)
%42.70.-a% (Optical materials)
%}

\noindent
\textbf{PROGRAM SUMMARY}

\begin{small}
\noindent
{\em Manuscript Title:} TIM, a ray-tracing program for forbidden ray optics \\
{\em Authors:} Dean Lambert, Alasdair C.\ Hamilton, George Constable, Harsh Snehanshu, Sharvil Talati, Johannes Courtial \\
{\em Program Title:} TIM \\
{\em Journal Reference:}                                      \\
  %Leave blank, supplied by Elsevier.
{\em Catalogue identifier:}                                   \\
  %Leave blank, supplied by Elsevier.
{\em Licensing provisions:} GNU GPL \\
  %enter "none" if CPC non-profit use license is sufficient.
{\em Programming language:} Java \\
{\em Computer:} Any computer capable of running the Java Virtual Machine (JVM) 1.6 \\
  %Computer(s) for which program has been designed.
{\em Operating system:}  Any; developed under Mac OS X Version 10.6 \\
  %Operating system(s) for which program has been designed.
{\em RAM:} typically 145 MB (interactive version running under Mac OS X Version 10.6) \\
  %RAM in bytes required to execute program with typical data.
{\em Keywords:} ray tracing, geometrical optics, METATOYs \\
  % Please give some freely chosen keywords that we can use in a
  % cumulative keyword index.
{\em Classification:} 14 Graphics, 18 Optics \\
  %Classify using CPC Program Library Subject Index, see (
  % http://cpc.cs.qub.ac.uk/subjectIndex/SUBJECT_index.html)
  %e.g. 4.4 Feynman diagrams, 5 Computer Algebra.
{\em External routines/libraries:} JAMA \cite{JAMA-in-summary} (source code included) \\
  % Fill in if necessary, otherwise leave out.
{\em Nature of problem:} \\
  %Describe the nature of the problem here.
visualisation of scenes that include scene objects that create wave-optically forbidden light-ray fields \\
{\em Solution method:} ray tracing \\
{\em Unusual features:}\\
  %Describe any unusual features of the program/problem here.
specifically designed to visualise wave-optically forbidden light-ray fields;
can visualise ray trajectories;
can visualise geometric optic transformations;
can create anaglyphs (for viewing with coloured ``3D glasses'') and random-dot autostereograms of the scene;
integrable into web pages \\
{\em Running time:}\\
  %Give an indication of the typical running time here.
Problem-dependent; typically seconds for a simple scene \\

\end{small}

% \noindent \textit{No.\ of lines in distributed program, including test data, etc.:}  approx.\ 30,900 (incl.\ comments); the source code of JAMA is another approx.\ 3,300 lines

\section{Introduction}

\noindent
TIM was originally conceived as a tool to allow experimentation (although only in the computer) with novel optical components called METATOYs \cite{Hamilton-Courtial-2009} prior to building them (TIM is an acronym for \emph{The Interactive METATOY}).
Since then, it has developed into a much more general ray-tracing program, suitable for use in optics research (including, but not limited to, METATOYs research), but also for simply playing with.

\begin{figure}
\begin{center} \includegraphics[width=\picturewidth]{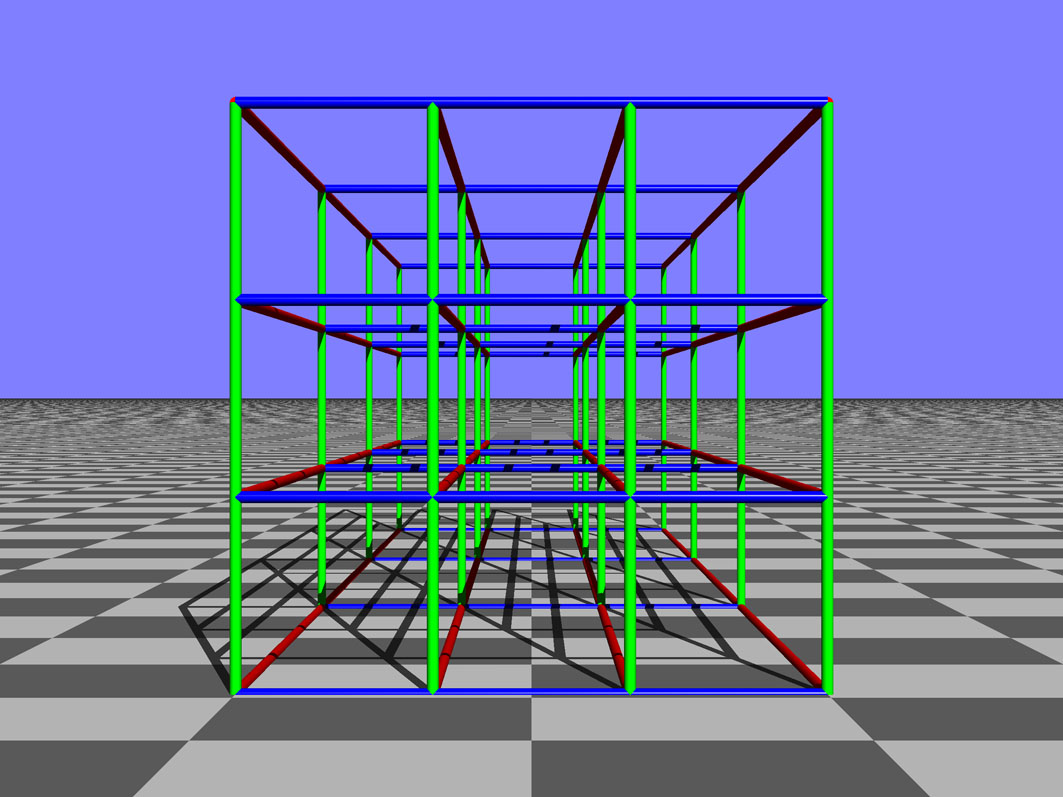} \end{center}
\begin{center} \includegraphics[width=\picturewidth]{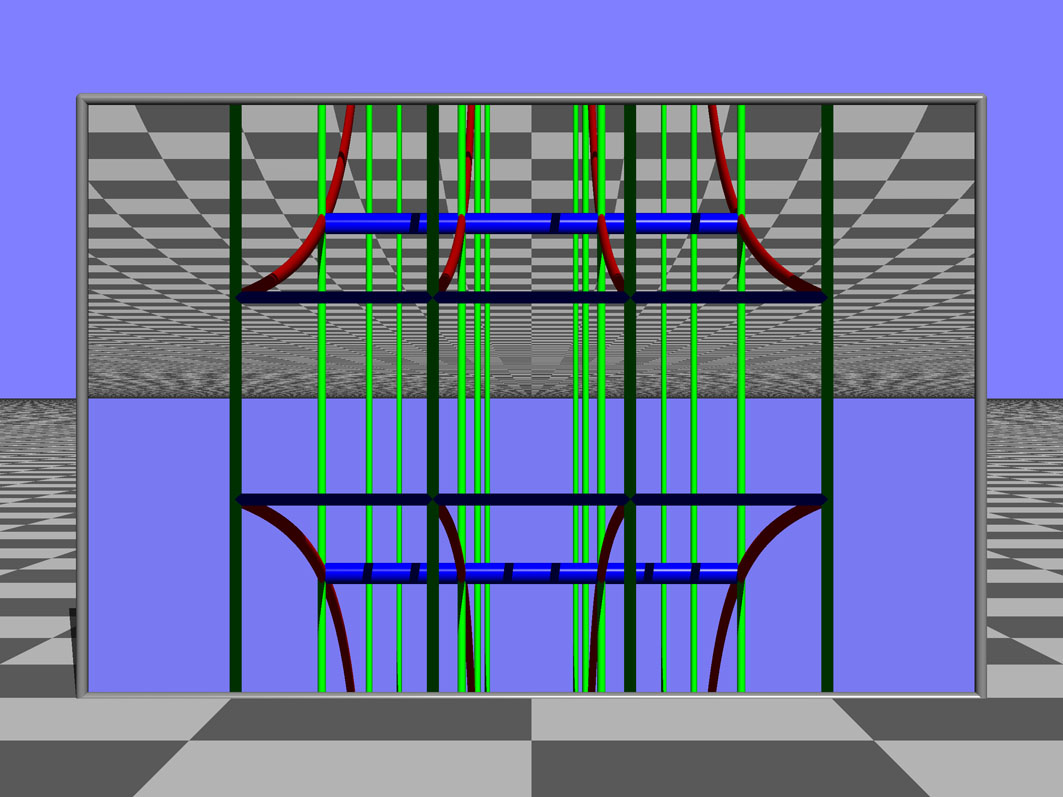} \end{center}
% \begin{center} \includegraphics[width=\picturewidth]{VerticalFlippingWindow3D.jpg} \end{center}
\caption{\label{vertical-flip-window-figure}Simulated view of a cylinder lattice, seen on its own (top) and through a window that changes the sign of the vertical light-ray-direction components~(bottom).
% framed rectangle, standard other than that it's centred at (0, 0, 8), flipping vertical component (0 degrees w.r.t. standard direction), in front of standard cylinder lattice (with 4 cylinders each in x and y direction)
}
\end{figure}

METATOYs are surfaces that appear to change the direction of light in ways that often result in wave-optically forbidden light-ray fields \cite{Courtial-et-al-2011}.
Of course, METATOYs cannot \emph{actually} create wave-optically forbidden light-ray fields; what they do is create light-ray fields that are visually almost indistinguishable from the forbidden fields.
% Wave-optically, the light-ray fields are forbidden because they have a fractional phase-vortex charge everywhere, and METATOYs concentrate those charges along lines.
Of particular interest to us are surfaces that perform a generalisation of refraction:
they change the direction of transmitted light rays according to laws that can be much more general than Snell's law.
In TIM, such generalised refraction can be described in terms of a complex representation, which is explained in section \ref{surface-property-section}.
The ability to handle very general surface properties, which enables the visualisation of scenes that include objects with METATOY surfaces (Fig.\ \ref{vertical-flip-window-figure}), is TIM's key speciality.

\begin{figure}
\begin{center} \includegraphics[width=\picturewidth]{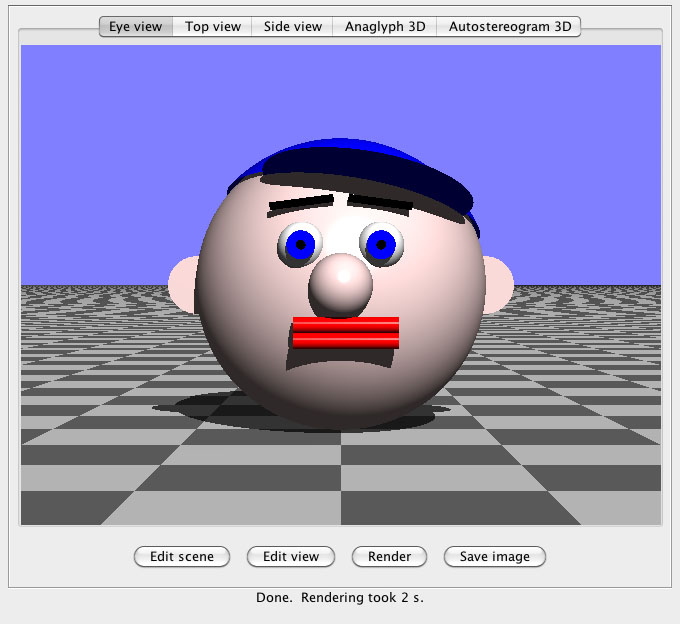} \end{center}
\caption{\label{TIM-window-figure}TIM, running as an interactive Java application on an Intel MacBook.
The central image is the rendered view of the default scene.
%, which includes three reflective spheres seen through a window that rotates the local light-ray direction by $90^\circ$ around the local surface normal \cite{Hamilton-et-al-2009}.
The Java applet version looks identical, apart from the ``Save image'' button being absent in the Java applet version because of security restrictions.}
\end{figure}

But TIM has other specialities, which support different aspects of our research.  TIM can
\begin{itemize}
\item simulate photos taken with cameras that can focus on almost arbitrary surfaces (section~\ref{arbitrary-focus-surface-camera});
\item simulate surfaces that ``teleport'' light rays to corresponding positions on other surfaces, from where the light rays then continue (section~\ref{teleporting-section});
\item visualise the trajectories of individual light rays (section~\ref{light-ray-visualisation-section});
\item render scenes as anaglyphs for viewing with red/cyan anaglyph glasses (section~\ref{anaglyph-section});
\item render scenes as random-dot autostereograms (section~\ref{autostereogram-section}).
\end{itemize}

There is one more speciality: TIM can be run as an interactive Java applet (Fig.\ \ref{TIM-window-figure}), which can easily be embedded in internet pages\footnote{An example can be found at \url{http://www.physics.gla.ac.uk/Optics/play/TIM/}.}.
We use this capability to disseminate our research over the internet, in a manner that invites playful exploration.
The use of this interactive version of TIM is described in a user guide~\cite{Lambert-et-al-2011}.

Sometimes our research requires capabilities which are not yet built into the interactive version of TIM.
This then requires modification of the source code, and sometimes the modifications become part of the interactive version of TIM.
As TIM is open-source software, in principle everybody can do this.
The aim of this paper is to encourage this:  to invite others to play with the interactive version of TIM, and to entice them to download and modify TIM's source code.
The paper contains several appendices aimed at facilitating the source-code modification by outlining
the implementation of ray tracing, which forms the core of TIM's source code (\ref{ray-tracing-appendix});
the overall structure of TIM's source code (\ref{structure-appendix});
and how to perform a number of code-modification tasks, including rendering a specific scene by modifying the source code for the default non-interactive TIM Java application (\ref{non-interactive-TIM-appendix}),
adding a new class of scene object (\ref{new-scene-object-appendix}),
and adding a new class of surface property (\ref{add-surface-property-appendix}).

%mathematics of finding intersections at \url{http://www.cl.cam.ac.uk/teaching/1999/AGraphHCI/SMAG/node2.html}

\section{\label{surface-property-section}Complex representation of generalised refraction}

\noindent
TIM exists because we wanted to see scenes that include METATOYs.
Surface properties describing METATOYs therefore play a key role in TIM.
TIM describes almost all METATOY surface properties in terms of a complex representation introduced in Ref.\ \cite{Sundar-et-al-2009}.

\begin{figure}
\begin{center} \includegraphics{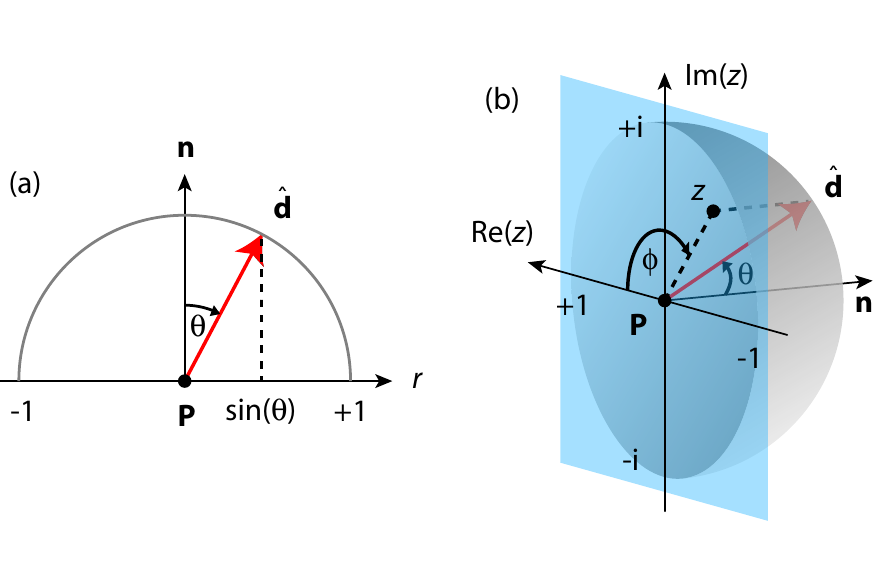} \end{center}
\caption{\label{complex-projection-figure}Representations of the normalised light-ray direction $\hat{\mathbf{d}}$ at a point $\mathbf{P}$ on a surface.
(a)~Orthographic projection onto a real axis tangential to the surface at $\mathbf{P}$ and with its origin at $\mathbf{P}$.
The real axis lies in the plane of incidence, which is the plane through $\mathbf{P}$ that is spanned by the local surface normal, $\mathbf{n}$, and the normalised light-ray direction, $\hat{\mathbf{d}}$.
(b)~ Orthographic projection into a complex plane tangential to the surface at $\mathbf{P}$ and with its origin at $\mathbf{P}$.}
\end{figure}

The complex representation is itself an extension of the way ray direction is represented in Snell's law,
\begin{equation}
n \sin \theta = n^\prime \sin \theta^\prime,
\end{equation}
where $\theta$ is the angle between the incident light-ray direction and the surface normal at the point $\mathbf{P}$ where the ray intersects the surface,
$\theta^\prime$ is the angle between the outgoing light-ray direction and the surface normal at $\mathbf{P}$,
and $n$ and $n^\prime$ are the refractive indices that meet at $\mathbf{P}$.
Each sine of an angle with the local surface normal can be interpreted as the number at the orthographic projection of the tip of the unit vector $\hat{\mathbf{d}}$ representing the corresponding light-ray direction onto a real axis that is tangential to the surface at $\textbf{P}$, has its origin located at $\textbf{P}$, and is lying in the plane of incidence (the plane spanned by the incident light-ray direction and the surface normal); this is shown in Fig.\ \ref{complex-projection-figure}(a).
Snell's law can then be interpreted as a simple multiplication of the projection of the light-ray direction by a factor $n / n^\prime$.

We now replace the real axis with a complex plane (Argand plane), again tangential to the surface at $\mathbf{P}$ and with its origin at $\mathbf{P}$.
Light-ray direction can then be described in terms of the complex number $z$ at the orthographic projection of the tip of the unit light-ray-direction vector $\hat{\mathbf{d}}$ into this complex plane (Fig.\ \ref{complex-projection-figure}(b)).
In this representation, Snell's law is still described by a simple multiplication of $z$ by a factor $n / n^\prime$.
That the outgoing light ray also lies in the plane of incidence is explicitly contained in this formulation, but not in Snell's law \cite{Sundar-et-al-2009}.
Rotation of the light-ray direction by an angle $\alpha$ around the local surface normal \cite{Hamilton-et-al-2009} is described by multiplication of $z$ by a factor $\exp(\rmi \alpha)$ \cite{Sundar-et-al-2009}.
Other light-ray-direction mappings can be described by other complex mappings $z \rightarrow z^\prime(z)$; the visual effects due to such mappings are investigated in more detail elsewhere~\cite{Constable-et-al-2011}.

\section{\label{arbitrary-focus-surface-camera}A camera that can focus on almost arbitrary surfaces}

\noindent
TIM has the ability to simulate photos taken with a camera that can focus on almost arbitrary surfaces.
In this section we explain how TIM simulates such a camera.

Focussing matters only in cameras with a non-zero aperture size.
(In photos taken by cameras with a point-like aperture --- pinhole cameras --- everything is imaged as sharply as diffractions permits.)
TIM simulates a camera with a finite-size aperture by backward-tracing, starting from each pixel, a number of light rays which have passed through different points on the aperture, and averaging the resulting colours.
The points on the aperture through which the backwards-traced light rays emerge are randomly chosen.
Which direction these light rays have as they leave the aperture is determined by the imaging properties of the lens:  the light rays originate at the position of a particular pixel, and so they have to leave the lens in the direction of the pixel's \emph{image} formed by the lens.

In a real lens, the images of all detector pixels lie, to a good approximation, in a plane.
By allowing the images of the detector pixels to lie on a much more general surface, the \emph{focus surface}, TIM simulates a camera which focusses on this focus surface.

The focus surface is defined as follows.
% TIM uses mechanisms already in place to perform ray tracing to define the focus surface.
The camera has associated with it a number of scene objects that define the so-called \emph{focus scene}, a scene in addition to the scene TIM renders.
The focus surface is then defined as those points in the focus scene visible to an observer placed at the centre of the aperture.

\begin{figure}
\begin{center} \includegraphics{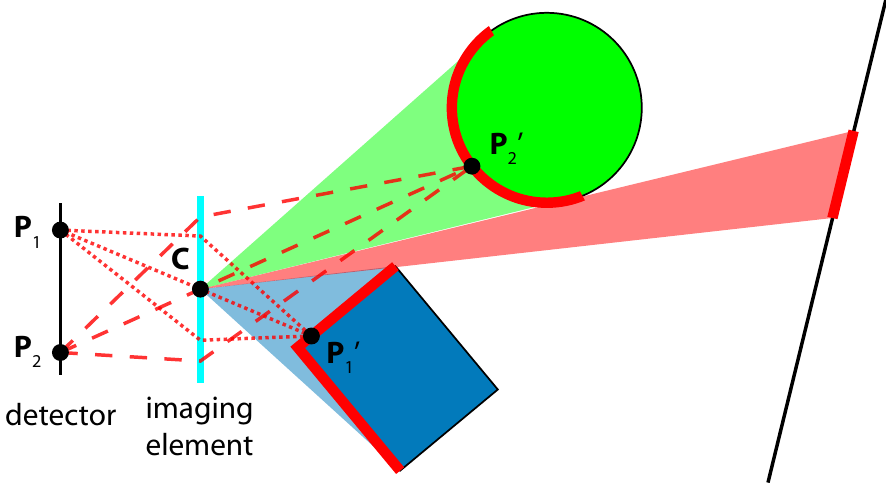} \end{center}
\caption{\label{focus-surface-figure}Construction of the focus surface and the position of the image of any detector pixel.
The focus surface (thick red line) consists of those parts of the focus scene visible from the centre $\mathbf{C}$ of the aperture of the imaging element.
The position of the image of any particular detector pixel $\mathbf{P}$ lies on the intersection between the focus surface and the continuation of the straight line between the pixel position and the point $\mathbf{C}$.
The figure shows the positions of the images of two pixels, $\mathbf{P}_1$ and $\mathbf{P}_2$.
In the example shown here, the focus scene consists of three objects:  a circle, a rectangle, and a line.}
\end{figure}

In the case of a thin lens, the image of any point, and specifically any detector pixel, lies somewhere along the continuation of the straight line from the point to the centre of the lens.
We generalise this here such that, in TIM's camera that focusses on non-planar surfaces, the image of any detector pixel lies somewhere along the continuation of the straight line from the pixel position to the centre of the aperture.
Fig.\ \ref{focus-surface-figure} illustrates this.

Conveniently, the position of the image of a particular pixel at position $\mathbf{P}$ can be found using functionality already built into TIM, namely its capability to find the closest intersection between an arbitrary light ray and a group of scene objects, one of the key capabilities for ray tracing:
all that is required is finding the closest intersection between the focus scene and a ray that starts from the centre of the aperture, $\mathbf{C}$, with a direction given by $(\mathbf{C} - \mathbf{P})$, the direction from the detector pixel to the centre of the aperture.

\begin{figure}
\begin{center} \includegraphics[width=\picturewidth]{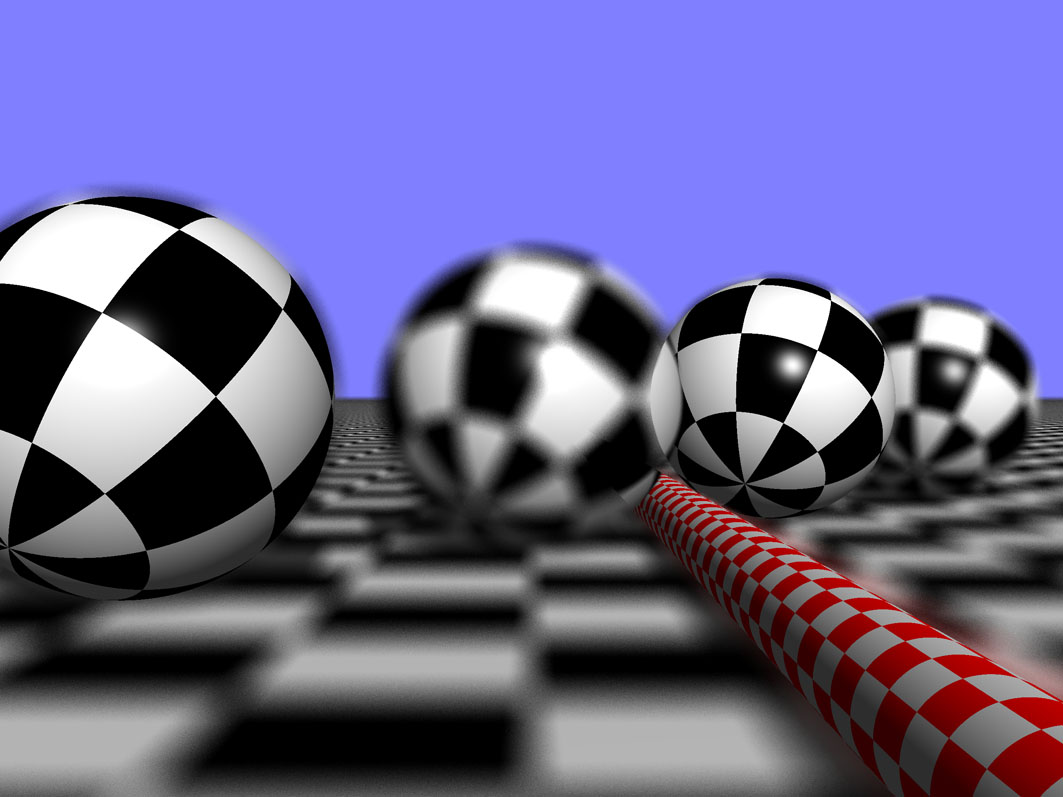} \end{center}
\caption{\label{blur-example-figure}Example of a scene rendered with a non-planar focus surface.
The focus scene consists of two of the four spheres, the chequered cylinder, and a plane in the far distance.}
\end{figure}

Fig.\ \ref{blur-example-figure} shows an example of a scene rendered for a non-planar focus surface.
The focus scene consists of a few --- but not all --- of the scene objects in the scene, and a distant plane.
The objects that are part of the focus scene can clearly be seen to be rendered sharply; those that are not are blurred.

%very closely related to some work on light-field cameras about focussing on inclined planes (in the case of a light-field camera this happens after a photo has been taken with a special camera that registers not only the direction from which light arrives at \emph{one} point, the pinhole position in the case of a pinhole camera)
%
%explain how it's possible to just focus on all the objects in the scene (with the right button), but also remark on the fact that this doesn't necessarily make everything sharp, like for example when a METATOY window is in the foreground (then the window will be in focus).

\section{\label{teleporting-section}Teleporting surface property}

\noindent
An ideal lens takes the field in one plane and creates an image of this field in another plane.
The image is stretched in the transverse directions, but not otherwise distorted.

Sometimes it is desirable to create an image that is distorted according to a specific mapping between the coordinates in the object and image planes.
This is called a geometrical optical transformation, and it can be approximated holographically \cite{Bryngdahl-1974}.
Our own interest in geometrical optical transformations stems from the application of a polar-to-Cartesian transformation between two planes to the detection of optical angular momentum~\cite{Berkhout-et-al-2010}.

\begin{figure}
\begin{center} \includegraphics[width=\picturewidth]{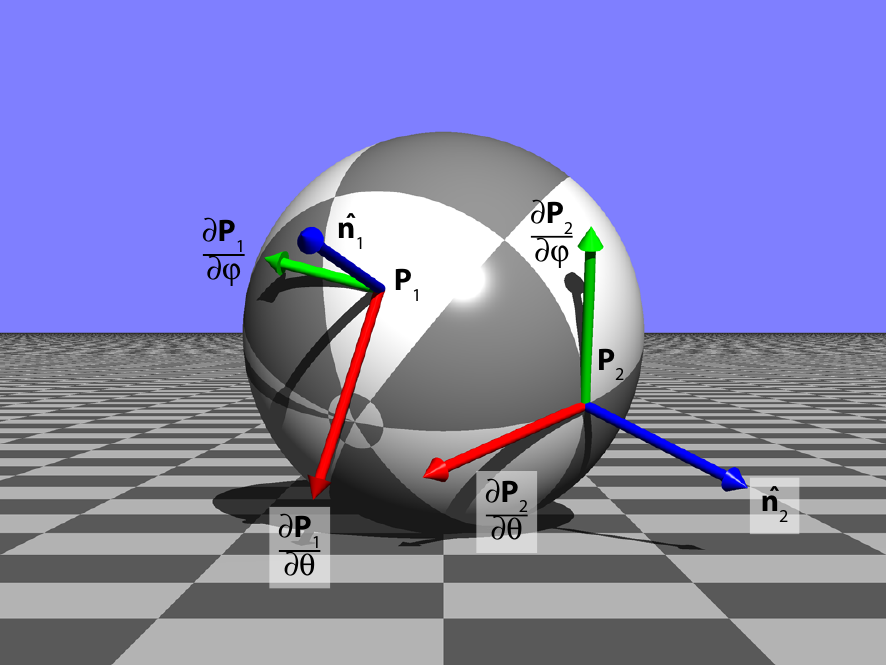} \end{center}
\caption{\label{parametrisation-figure}Parametrisation of a scene object.
Each point on the surface is described by a pair of surface coordinates, in the picture the spherical coordinates $\theta$ and $\phi$, defined with respect to an arbitrary zenith direction (here $(0.408, 0.408, 0.816)$) and direction of the azimuth axis (here $(0.913, -0.183, -0.365)$).
This parametrisation of the surface has been indicated by covering it in a chequerboard pattern with tiles of side length 1 in both $\theta$ and $\phi$.
The local surface-coordinate axes, $\hat{\mathbf{\theta}}_i = \partial \textbf{P}_i / \partial \theta$ and $\hat{\mathbf{\phi}}_i = \partial \textbf{P}_i / \partial \phi$, together with the local surface normals, $\hat{\mathbf{n}}_i$, are shown for two points, $\mathbf{P}_i$ ($i = 1, 2$).
% These vectors are shown for two points, namely for $(\theta, \phi) = (2.5, 1.5)$ and for $(\theta, \phi) = (2,0)$.
The sphere has radius 1 and is centred at $(0, 0, 10)$.}
\end{figure}

For geometrical optical transformations, coordinates are clearly important.
Many of TIM's scene objects have associated with them a two-dimensional coordinate system that parametrises their surface, i.e.\ each point $\mathbf{P}$ on the scene object's surface is described by a pair of associated surface coordinates, $c_1$ and $c_2$.
For example, positions on a plane are described by Cartesian surface coordinates;
positions on a circular disc are described by their polar coordinates $r$ (the distance from the centre) and $\phi$ (the azimuthal angle);
positions on a sphere are described by their spherical polar coordinates $\theta$ (the polar angle) and $\phi$ (the azimuthal angle).
Scene objects parametrised in this way can also calculate the local surface-coordinate axes for any point on the surface.
These are the vectors
\begin{equation}
\hat{\mathbf{c}}_1 = \frac{\partial \mathbf{P}}{\partial c_1}, \quad
\hat{\mathbf{c}}_2 = \frac{\partial \mathbf{P}}{\partial c_2};
\end{equation}
they respectively point in the direction in which the corresponding surface coordinate changes at the point $\mathbf{P}$, and their respective length indicates the distance on the surface over which the corresponding surface coordinate changes by 1.
Fig.\ \ref{parametrisation-figure} shows the local surface-coordinate axes for two positions on a sphere.
The surface coordinates and surface-coordinate axes, respectively, play a key role in the calculation of the starting point and direction of the continuation of the incident light ray.

TIM can simulate geometrical optical transformations using an unusual surface property.
In TIM's implementation of ray tracing (see \ref{ray-tracing-appendix}), surface properties are responsible for returning the colour of light leaving a specific point on the surface in a specific direction.
Finding this colour often requires further tracing of the incident ray, for example in the case of a specularly reflective surface, where the continuation of the ray leaves the same point on the surface with a new direction given by the law of reflection.
Geometrical optical transformations can therefore be implemented in the form of a surface property that continues tracing the incident light ray, starting from a transformed position and with a suitably transformed direction.

\begin{figure}
% \begin{center} \includegraphics[width=\picturewidth]{cylinderLattice8to18.jpg} \end{center}
\begin{center} \includegraphics[width=\picturewidth]{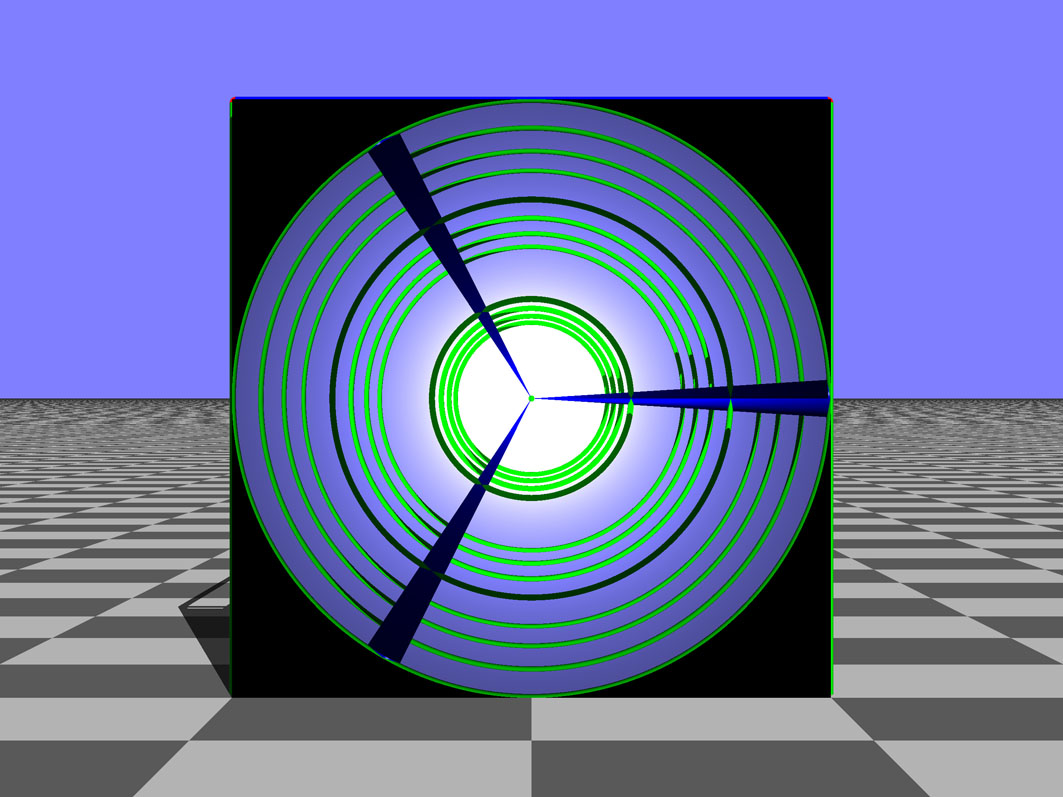} \end{center}
\caption{\label{Cartesian-2-polar-converter-figure}Simulated view through a Cartesian-to-polar converter.
In the example shown here, the converter has been placed immediately in front of the cylinder lattice shown in Fig.\ \ref{vertical-flip-window-figure}.
The converter distorts vertical cylinders (green) into circles centred on the origin, and horizontal cylinders (blue) into radial lines.
A brightening of the central region due to area elements being transformed to a different size (see Eqn~(\ref{intensity-change-equation})) is clearly visible.
The converter consists of a circular disc (the origin object), parametrised in terms of distance from the centre and azimuthal angle $\phi$, % scaled such that they take values between $0$ and $1$ on the surface.
with a teleporting surface whose target object is a black square immediately behind the disc, parametrised in terms of Cartesian coordinates.
All coordinates are scaled to range from 0 to 1.
% The range of the parameters describing both objects is chosen such that  the surface of the disc is mapped onto the
% disc centred at $(0, 0, 8)$, radius 1.
%cylinder lattice: $z = 8.1$ to $18.1$.
}
\end{figure}

TIM's teleporting surface property does precisely this.
A teleporting surface has associated with it a destination object; note that the link is one-way.
We call the object with the teleporting surface the origin object.
The mapping from the position $\mathbf{P}$ where the incident light ray intersects the origin object to the position where its continuation leaves the destination object's surface, $\mathbf{P}^\prime$, is defined in terms of the surface-coordinate systems (Fig.\ \ref{parametrisation-figure}) associated with the two objects' surfaces.
Specifically, if the position $\mathbf{P}$ where the incident light ray intersects the origin object's surface is described by some values of its surface coordinates, then the point $\mathbf{P}^\prime$ where the ray's continuation leaves the destination object's surface is the point at which the destination object's surface coordinates take those same values.
For example, if a planar origin object is parametrised in terms of polar surface coordinates $r$ and $\phi$, and the destination object associated with its teleporting surface property is also planar and parametrised in terms of Cartesian surface coordinates, then this setup is a polar-to-Cartesian converter for backwards-propagating light rays, which means it is a Cartesian-to-polar converter for forwards-propagating light rays.
Fig.\ \ref{Cartesian-2-polar-converter-figure} shows a three-dimensional (3D) lattice of cylinders seen through such a setup.
It can clearly be seen that lines of constant $z$ and $x$ value (i.e.\ vertical lines) in the cylinder lattice become lines of constant radius (i.e.\ circles) when seen through the converter, and that lines of  constant $z$ and $y$ value (horizontal lines) become lines of constant azimuthal angle (spokes).

The corresponding mapping of the light-ray direction is based on wave-optical considerations.
Specifically, we assume that phase and intensity of the (scalar) optical field on the teleporting surface gets mapped onto the destination object's surface.
In the ray-optics limit of wave optics \cite{Landau-Lifshitz-II-light-rays}, which is arguably appropriate here, the light-ray direction is proportional to the direction of the gradient of the phase. %, $\phi(x, y, z)$.
The wave's local phase gradient then defines the new light-ray direction.
In TIM, light-ray direction is normalised and therefore naturally interpreted as normalised phase gradient.

The components of the phase gradient are, of course, the rate of change of phase in the corresponding directions.
Consider the two coordinate systems that describe the surfaces of the origin object and of the destination object.
Now consider the component of the incident light-ray direction in the direction of the origin object's first surface coordinate, $c_1$.
A value $g_1$ means that the phase changes locally at a rate of $g_1$ full $2 \pi$ phase cycles over the distance on the surface in which $c_1$ changes by 1.
On the destination object, the phase then changes locally (at the point $\mathbf{P}^\prime$ where the light ray continues) at a rate of $g_1$ full $2 \pi$ phase cycles over the distance in which the destination object's first surface coordinate, $c_1^\prime$, changes by 1.
The ratio of the phase gradients in the direction of the first surface coordinate, which is the ratio of the light-ray components in the direction of the first surface coordinates, is therefore given by the ratio of the distances over which the first surface coordinate changes by 1 in the origin object and in the destination object.
These distances are given by length of surface-coordinate axes, $|\hat{\mathbf{c}}_1| = |\partial \mathbf{P} / \partial c_1|$ and $|\hat{\mathbf{c}}^\prime_1| = |\partial \mathbf{P}^\prime / \partial c_1|$.
A similar argument can be made for the second surface coordinate, $c_2$.
The components of the direction of the continuation of the light ray in the directions of the destination object's first and second surface coordinates, $d^\prime_1$ and $d^\prime_2$, are then
\begin{equation}
d^\prime_1 = \frac{| \hat{\mathbf{c}}_1 |}{| \hat{\mathbf{c}}^\prime_1 |} d_1, \quad
d^\prime_2 = \frac{| \hat{\mathbf{c}}_2 |}{| \hat{\mathbf{c}}^\prime_2 |} d_2,
\end{equation}
where  $d_1$ and $d_2$ are the components of the direction of the incident light ray in the directions of the first and second surface coordinates of the origin object.

The component of the light-ray-direction vector in the direction of the surface normal is chosen such that the length of the light-ray-direction vector remains unchanged.
This correctly represents the case of the continuation of the ray travelling in a medium with the same refractive index as the medium in which the incident ray was travelling.

One further consideration is a concomitant change in brightness.
TIM assumes that no power is lost during teleportation, so the power entering an area element $\rmd A$ on the surface of the origin object at position $P$ is the same as that exiting the corresponding area element $\rmd A^\prime$ on the surface of the destination object at $P^\prime$.
The light intensities (power per area) at $P$ and $P^\prime$, $I$ and $I^\prime$, are then given by the equation $I \rmd A = I^\prime \rmd A^\prime$.
The ratio of the area elements is given by
\begin{equation}
\frac{\rmd A^\prime}{\rmd A} =
\frac{\left| \hat{\mathbf{c}}_1^\prime \times \hat{\mathbf{c}}_2^\prime \right|}
{\left| \hat{\mathbf{c}}_1 \times \hat{\mathbf{c}}_2 \right|},
\end{equation}
and so
\begin{equation}
I = I^\prime \frac{\left| \hat{\mathbf{c}}_1^\prime \times \hat{\mathbf{c}}_2^\prime \right|}
{\left| \hat{\mathbf{c}}_1 \times \hat{\mathbf{c}}_2 \right|}.
\label{intensity-change-equation}
\end{equation}

It is worth noting that the teleporting surface property can be used for purposes other than implementing geometrical optical transformations.
For example, consider a planar origin object placed immediately in front of the camera and a planar target object placed elsewhere, both parametrised in terms of Cartesian coordinate systems.
If the scale of the two coordinate systems is the same, i.e.\ if the surface-coordinate axes on the origin and target objects are of the same length, then the effect is a simple change of camera position and viewing direction.

\section{\label{light-ray-visualisation-section}Visualisation of light-ray trajectories}

\begin{figure}
\begin{center} \includegraphics[width=\picturewidth]{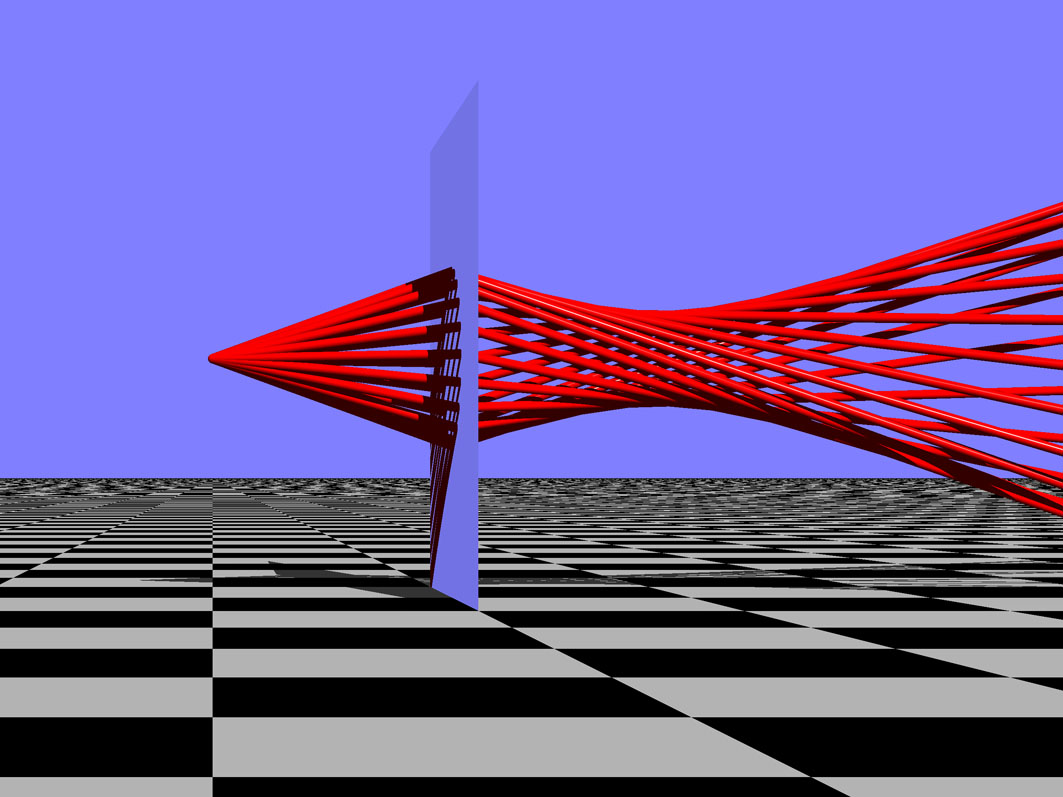} \end{center}
\caption{\label{light-ray-cone-figure}Visualisation of light-ray trajectories.
A cone of light-ray trajectories originating from a point in front of a window that rotates the light-ray direction by $150^\circ$ around the local window normal is converted into a twisted bundle of rays.
% cone apex $(x,y,z)=(-1,0,10)$
}
\end{figure}

\noindent
One of TIM's capabilities, namely the visualisation of light-ray trajectories, can be very helpful in understanding the effect of optical components.
Fig.\ \ref{light-ray-cone-figure} shows a cone of light rays being traced through a window that rotates the local light-ray direction through $90^\circ$, showing, for example, that such a window would not create an image of a point light source at the apex of the cone.

TIM visualises the trajectories of specific light rays in three steps:
\begin{enumerate}
\item trace the light rays, keeping a list of the points where each light ray intersects a scene object;
\item for each segment of the above light-ray trajectory, i.e.\ between each pair of neighbouring intersection points, add a cylinder to the scene;
\item visualise the scene.
\end{enumerate}

The first point uses the ray-tracing methods already built into TIM, but those methods needed to be extended slightly to keep track of ray trajectories.
This requires the ability to deal with rays branching, which occurs whenever a ray encounters an object with multiple surface properties that require further ray tracing, such as a partially transmissive mirror.
In TIM, whenever a ray hits a surface that requires further ray tracing, a new ray is created, added to the list of the ray's branches, and traced.
A light ray's full trajectory is then stored in the form of a list of positions where the main branch intersects scene objects, and the list of the branches.

\section{\label{anaglyph-section}Anaglyphs}

\noindent
Parallax causes a scene to look different when viewed from two different positions.
This is called binocular disparity; the process of deriving depth perception from the two different views the two eyes receive is called stereopsis \cite{Palmer-1999-stereopsis}.
In anaglyph images \cite{Wikipedia-AnaglyphImage}, the two views are superposed, but in different colours.
When viewed through suitable anaglyph glasses, i.e.\ a different colour filter in front of each eye which in the simplest case completely filters out the view intended for the other eye, different images can be presented to the eyes, and stereopsis can lead to depth perception.

\begin{figure}
% \begin{center} \includegraphics[width=\picturewidth]{CylinderLattice.jpg} \end{center}
\begin{center} \includegraphics[width=\picturewidth]{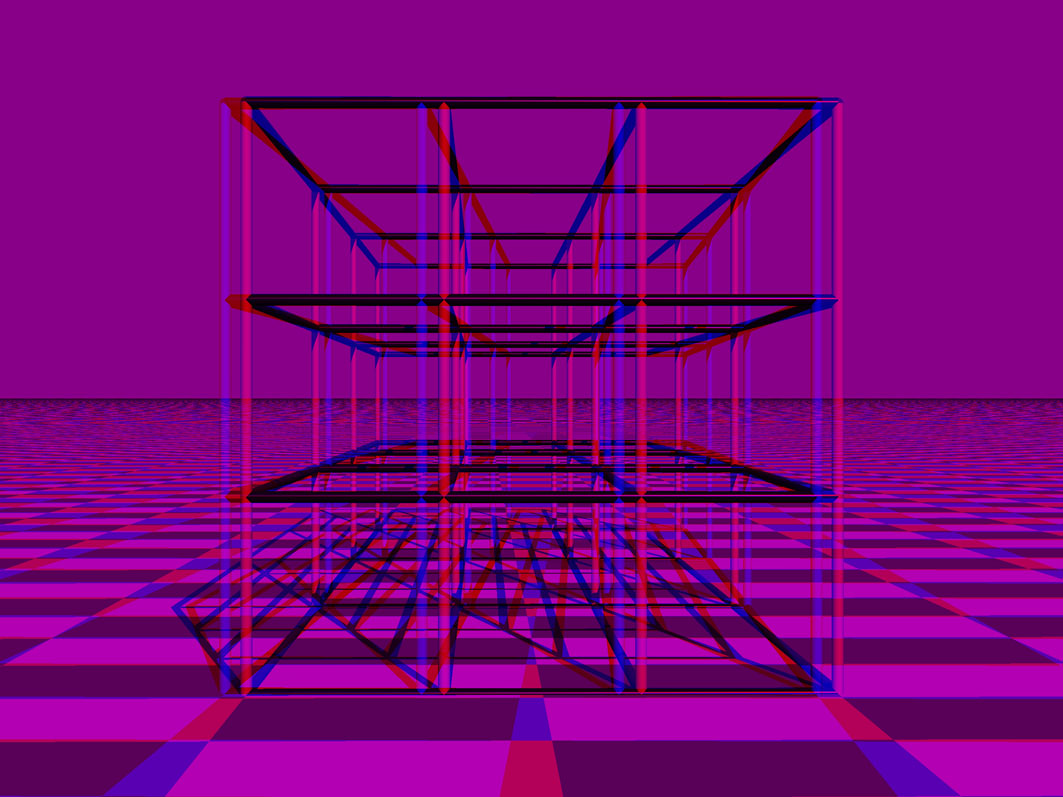} \end{center}
\begin{center} \includegraphics[width=\picturewidth]{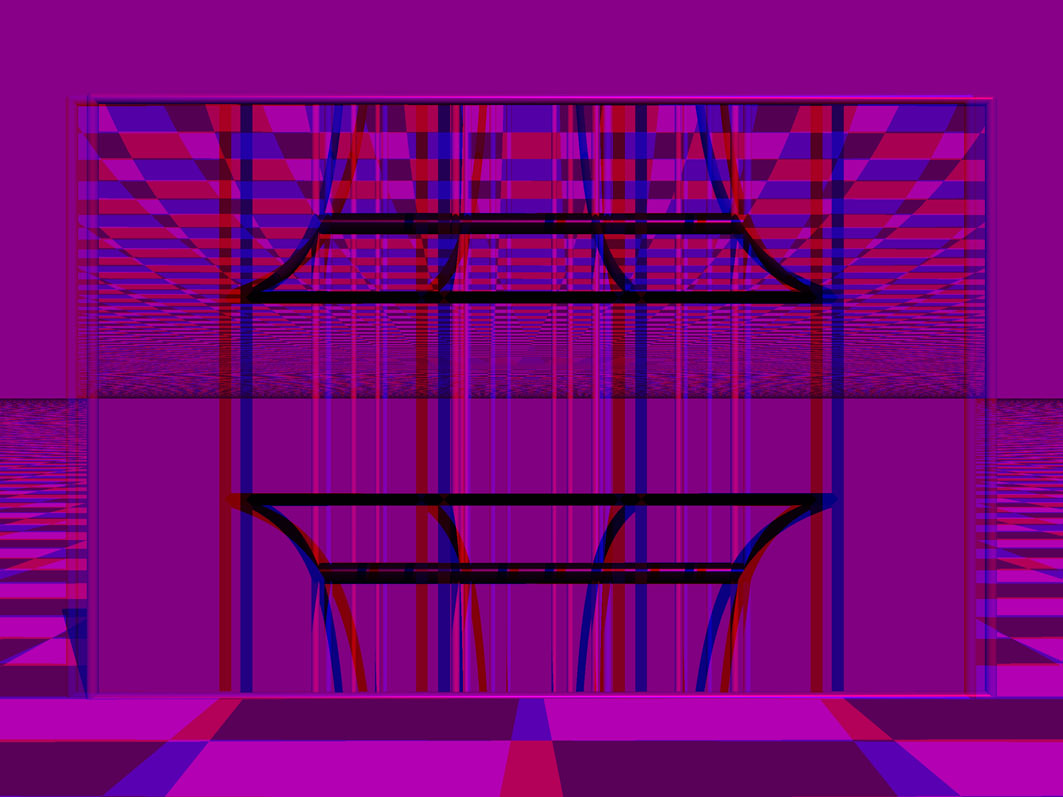} \end{center}
\caption{\label{anaglyph-images-figure}Anaglyph versions of the images in Fig.\ \ref{vertical-flip-window-figure}, which show a cylinder lattice au naturel~(top) and when seen through a window that inverts the vertical ray-direction component~(bottom).
% The parameters for the 3D view are look-at-point (0, 0, 10), eye separation 0.4, aperture size pinhole, anti-aliasing quality great
}
\end{figure}

%\begin{figure}
%\begin{center} \includegraphics[width=\picturewidth]{VerticalFlippingWindow.jpg} \end{center}
%\begin{center} \includegraphics[width=\picturewidth]{VerticalFlippingWindow3D.jpg} \end{center}
%\caption{framed rectangle, standard other than that it's centred at (0, 0, 8), flipping vertical component (0 degrees w.r.t. standard direction), in front of standard cylinder lattice (with 4 cylinders each in x and y direction)
%3D view, look-at-point (0, 0, 10), eye separation 0.4, aperture size pinhole, anti-aliasing quality great.
%As far as seeing with two eyes is concerned, this is a perfect, but distorted, image.}
%\end{figure}

\begin{figure}
% \begin{center} \includegraphics[width=\picturewidth]{anaglyphRedBlueDemo.jpg} \end{center}
\begin{center} \includegraphics[width=\picturewidth]{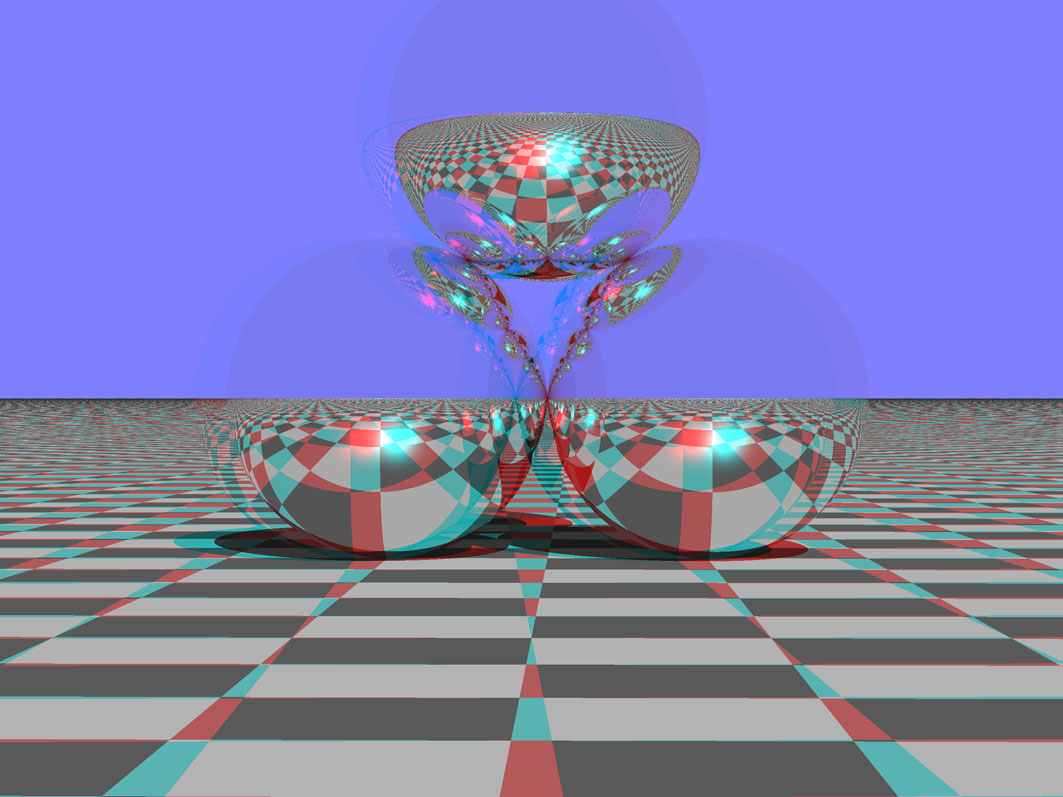} \end{center}
\caption{\label{anaglyph-colour-figure}Example of a colour anaglyph image.
In addition to the chequerboard floor, the scene contains three reflective spheres.
% is the default scene with the ray-rotating window and its frame removed.
}
\end{figure}

TIM can create anaglyph images intended for viewing with red/cyan anaglyph glasses.
Two images are calculated for camera positions that differ by a sideways displacement.
These two images can then be turned into anaglyph images in two different ways:
\begin{enumerate}
\item The red component of the image is the luminosity of the left-eye image, the blue component is the luminosity of the right-eye image.
The resulting anaglyph has lost colour information.
\item Following ``recent simple practice'' \cite{Wikipedia-AnaglyphImage}, the blue and green components are removed from the image that corresponds to the left eye, the red component is removed from the right-eye image, and the two images are superposed.
The resulting anaglyph includes colour information, but does not work very well for objects of certain colours.
\end{enumerate}
Figures \ref{anaglyph-images-figure} and \ref{anaglyph-colour-figure} show examples of anaglyph images.

%display them in various ways, e.g.\
%\begin{itemize}
%\item red-blue anaglyphs;
%\item show one, then the other, then the first again, etc.;
%\item show both views side-by-side.
%\end{itemize}

\section{\label{autostereogram-section}Random-dot autostereograms}

\begin{figure}
\begin{center} \includegraphics[width=\picturewidth]{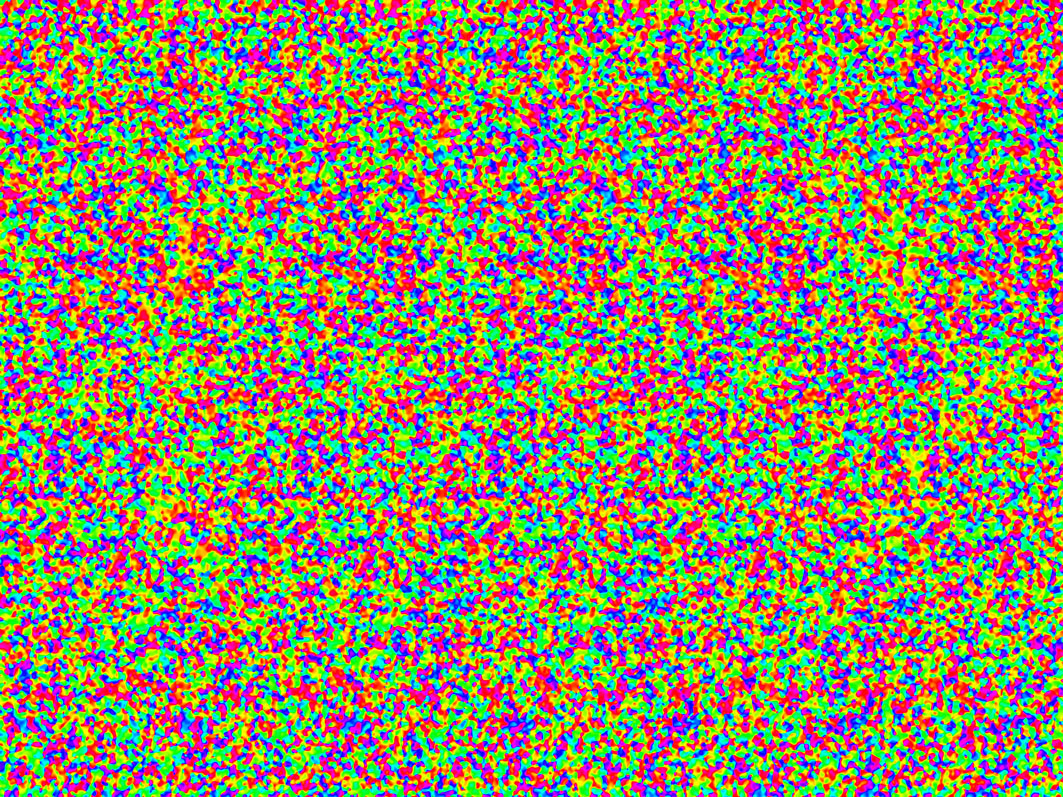} \end{center}
\caption{\label{autostereogram-example-figure}Random-dot autostereogram of TIM's default scene (Fig.\ \ref{TIM-window-figure}).
Note that the scene includes a plane behind Tim's head which can be seen in the background, making it easier to see the autostereogram.
% As the plane is transparent and does not throw a shadow, it is normally invisible in all other views.
}
\end{figure}

\noindent
TIM can render scenes as random-dot autostereograms \cite{Tyler-Clarke-1990}.
Fig.\ \ref{autostereogram-example-figure} shows an example of such a random-dot autostereogram created by TIM.

\begin{figure}
\begin{center} \includegraphics{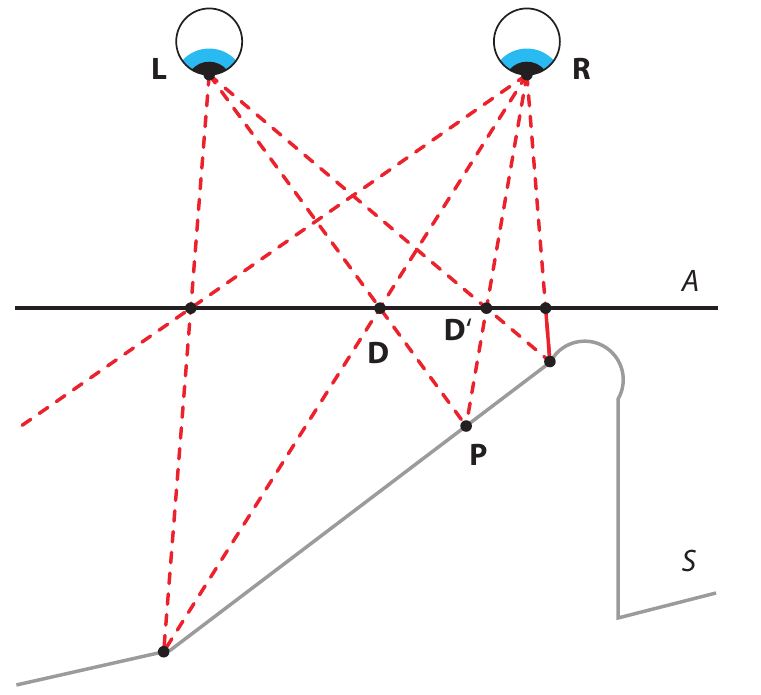} \end{center}
\caption{\label{random-dot-autostereogram-figure}Principle of random-dot autostereograms.
The eyes are located at the positions $\mathbf{L}$ and $\mathbf{R}$; the autostereogram is in the plane $A$; $S$ is the 3D surface represented by the autostereogram.}
\end{figure}

Autostereograms rely on the fact that two different patterned surfaces can look identical from different viewing positions.
In the case of standard autostereograms, the two different viewing positions are the positions of the observer's eyes, one of the surfaces is planar, and the pattern in that plane is the autostereogram of the other, three-dimensional, surface.
For the observer to perceive the visual illusion of seeing the three-dimensional surface when viewing the autostereogram requires the patterns to be sufficiently detailed.

Placing dots on the two surfaces such that the patterns are consistent with each other is perhaps the simplest way to construct autostereograms.
TIM uses the following algorithm.
We call the plane of the autostereogram $A$, and the three-dimensional surface $S$ (see Fig.\ \ref{random-dot-autostereogram-figure}).
\begin{enumerate}
\item Randomly pick a dot colour.
\item \label{pick-D} Randomly pick a position $\mathbf{D}$ in the plane $A$ and place a small dot of the picked dot colour there.
\item \label{find-P} Find the point $\mathbf{P}$ on $S$ that lies on the same line of sight as $\mathbf{D}$, as seen from the position of the left eye, $\mathbf{L}$.
For the surfaces $A$ and $S$ to look identical from the position of the left eye, $\mathbf{P}$ therefore has to have the same colour as $\mathbf{D}$, namely the dot colour picked above.
\item Find the point $\mathbf{D}^\prime$ on $A$ that lies on the same line of sight as $\mathbf{P}$, as seen from the position of the \emph{right} eye, $\mathbf{R}$.
By the same argument as above, the colour of this point must also be the dot colour picked above.
Therefore, place another small dot of the picked dot colour at $\mathbf{D}^\prime$.
\item \label{find-Ds} The previous two steps constructed, from the position $\mathbf{D}$ of one dot in the autostereogram, the position $\mathbf{D}^\prime$ of another dot that has to have the same colour.
Repeat the previous two steps to construct, from the position $\mathbf{D}^\prime$ of this other dot, the position of yet another dot that has to have the same colour.
Keep repeating until the position of the new dot lies outside the area of the autostereogram.
\item Steps \ref{find-P} to \ref{find-Ds} constructed, from the position $\mathbf{D}$ picked in step \ref{pick-D}, the positions of further dots that have to be of the same colour.
This was done by using the left eye's line of sight to construct a corresponding point on $S$, and then the right eye's line of sight to construct a point on $A$ corresponding to this new point.
Start again from the position picked in step \ref{pick-D}, $\mathbf{D}$, and repeat steps \ref{find-P} to \ref{find-Ds}, but swapping the role of the left eye and the right eye.
In other words, now use the \emph{right} eye's line of sight to construct a point on $S$ that corresponds to a dot position in $A$, and then use the \emph{left} eye's line of sight to construct a point on $A$ corresponding to this point on $S$, which is then the position of a new dot.
\end{enumerate}

Two details in TIM's algorithm are perhaps worth noting.
\begin{enumerate}
\item The dots placed by TIM in the autostereogram are not single pixels but Gaussians.
This means that their positions are not restricted to the pixel positions, so they can be placed ``between pixels''.
This is advantageous as restricting dot positions to pixel positions restricts the possible separations between dots, and therefore the depths the dot patterns can represent, which results in ``layered'' random-dot stereograms.
\item  Each dot TIM places in the autostereogram has a particular hue.
To calculate the colour of a particular pixel in the autostereogram, TIM calculates the weighted average of the hues of all the dots that intersect at the pixel.
Hue is the azimuthal angle $\phi$ of a colour's position in a suitable colour wheel (e.g.\ red $=0^\circ$, yellow $=60^\circ$, green $=120^\circ$, cyan $=180^\circ$, blue $=240^\circ$, purple $=300^\circ$). 
To form the weighted average of the hues due to all dots, TIM converts the hue of each dot into corresponding Cartesian coordinates (the coordinates corresponding to the hue $\phi_j$ of the $j$th dot, weighted by a factor $w_j$, are $x_j = w_j \cos(\phi_j)$ and $y_j = w_j \sin(\phi_j)$), adding up the $x$ and $y$ coordinates due to all hues, and calculating the azimuthal angle $\phi$ of the resulting $(x,y)$ position.
The weight of the hue of a particular dot is given by $w_j = \exp \left[-(r_j /\rho)^2 \right]$, where $r_j$ is the distance between the pixel and the centre of the $j$th dot, and $\rho$ is the dot radius.
This can be expressed in terms of complex numbers as
\begin{equation}
\phi = \arg \left\{ \sum_j \exp \left[-(r_j/\rho)^2 \right] \exp(\rmi \phi_j) \right\}.
\end{equation}
Finally, the pixel colour is found using the standard Java method for converting a colour represented in terms of its hue, saturation and brightness (HSB) into its red, green and blue (RGB) components.
Both saturation and brightness are set to their maximum value, 1.
\end{enumerate}

\section{Conclusions}

\noindent
TIM is a powerful raytracer with extensive capabilities, a number of them unique.
We use TIM in our research on METATOYs:
% for modelling light propagation through METATOY structures of interest;
for disseminating our research over the internet by inviting playful experimentation with METATOYs through TIM's interactive version;
and for conducting computer experiments with METATOYs ourselves.

Sometimes it is necessary to modify TIM's source code, which can be a daunting prospect.
This paper is intended to help others and ourselves doing this.
We hope it will entice other researchers to use TIM in their own work.

\section*{Acknowledgements}

\noindent
Thanks to The MathWorks and to NIST for making the source code of JAMA publicly available.

%\section{Object catalogue}
%
%\begin{figure}
%\caption{\label{sphere-figure}Sphere, radius 1, centred at $(0, 0, 10)$, tiled.
%In ``Top view'' and ``Side view'', the $z$ range is 4.
%Anti-aliasing quality ``Good''.}
%\end{figure}
%
%\begin{figure}
%\caption{Cylinder, radius 1, axis between $(-2, 0, 10)$ and $(+2, 0, 10)$, tiled.
%View parameters like in Fig.\ \ref{sphere-figure}.}
%\end{figure}

\appendix

\section{\label{ray-tracing-appendix}Implementation of ray tracing in TIM}

%Here we describe the key parts of TIM's optics engine, which is responsible for tracing rays.
%These are frequently encountered when modifying TIM's source code.

\noindent
TIM is written in Java \cite{Wikipedia-Java}, an object-orientated programming language \cite{Wikipedia-ObjectOrientedProgramming}.
As in all object-orientated languages, objects are instances of classes.
A class has associated methods and data (variables), which take on specific values in specific instances of the class, i.e.\ the objects.
A class can have subclasses, which inherit its methods and properties\footnote{Unless they prevent this specifically by explicitly overriding individual methods and/or properties.}; the former is called the latter's superclass.
In Java there are also interfaces, which are collections of methods.
If an object implements a specific interface, it has to implement all the interface's methods.

To trace a ray in TIM, a number of objects are interacting.
The following are the most important ones.
\begin{description}
\item[Rays.]  Each ray is an object of class \texttt{Ray}, which has a starting position and a normalised direction, both represented as 3D Cartesian vectors.

\item[Scene objects.]  Each scene object implements the interface \texttt{SceneObject}\footnote{Note that every Java class is a subclass of the \texttt{Object} class.}; % optics.raytrace.sceneObjects
most (but, for complicated reasons, not all) are instances of a subclass of \texttt{SceneObjectClass}, which implements the \texttt{SceneObject} interface.
Specifically, a scene object implements methods that calculate the intersection point between the scene object and any ray, and methods that calculate the colour of an intersection point when seen along the direction of a given ray and under specified illumination.
There are particularly fundamental scene objects, each typically representing a simple geometrical object (such as a sphere) with specific surface properties (such as reflective), which are instances of the \texttt{SceneObjectPrimitive} subclass of \texttt{SceneObjectClass}.
Another subclass of \texttt{SceneObjectClass} is \texttt{SceneObjectContainer}, which represents a collection of scene objects.
There are also more complicated scene objects that are neither \texttt{SceneObjectPrimitive}s nor \texttt{SceneObjectContainer}s, but ultimately every intersection with a \texttt{SceneObject} must be traceable to an intersection with a \texttt{SceneObjectPrimitive}.
%Each \texttt{SceneObject} is ultimately made up from \texttt{SceneObjectPrimitive}s; in the simplest case, it \emph{is} a \texttt{SceneObjectPrimitive}, but it can also be a collection of scene objects calle
%A number of classes implement the \texttt{SceneObject} interface, including \texttt{SceneObjectPrimitive}, which typically represents a simple geometrical object (such as a sphere) with specific surface properties (such as reflective);
%and \texttt{SceneObjectContainer}, which represents a collection of scene objects.
%Each \texttt{SceneObject} ultimately consists of \texttt{SceneObjectPrimitive}s, which 

\item[Surface properties.]
Perhaps the simplest surface property is a colour that is independent of any light sources (effectively a coloured glow), which is represented by the class \texttt{SurfaceColourLightSourceIndependent}.
The colour of non-glowing surfaces depends on illumination, and this is represented by the class \texttt{SurfaceColour}, in which the surface has separate diffuse and specular colours.
The diffuse component colours light due to Lambertian reflectance~\cite{Wikipedia-LambertianReflectance};
the specular components colours light that is specularly (or near-specularly) reflected~\cite{Wikipedia-PhongShading}.

There are also classes representing surfaces on which the ray does not end, but which change its direction.
Examples include mirror surfaces (\texttt{Reflective}) and refractive-index interfaces (\texttt{Refractive}).

Finally, there is currently one class (\texttt{Teleporting}) which continues tracing a light ray with a changed direction and from a new starting position.

All surface properties implement the interface \texttt{SurfaceProperty}.

%\item[Intersections between rays and scene objects.]  These are represented by objects of class \texttt{RaySceneObjectIntersection}.
%Such an object is simply a collection of the information describing the intersection between a ray and an object, such as the coordinates of the intersection point and the \texttt{SceneObjectPrimitive} object that was intersected.

\item[Light sources.]  Light sources are represented by instances of the class \texttt{LightSource}.
In TIM, two types of light source are implemented:  ambient light, represented by the class \texttt{AmbientLight}, and light sources with a specific 3D position which throw shadows and which can create highlights due to specular (or near-specular) reflection off surfaces of type \texttt{SurfaceColour}.
% Phong model

\end{description}

Ray-tracing software usually considers only those light rays that eventually enter the virtual camera.
They do this by tracing light rays \emph{backwards}, starting from each pixel in the camera's virtual detector chip.
What they try to do is establish the colour such a light ray would have if it was travelling in the opposite direction, i.e.\ the colour of the reverse ray, which is the colour an observer would see in the direction of the light ray.
More details can be found in Ref.\ \cite{Wikipedia-rayTracing}.

Backwards tracing a specific light ray in TIM proceeds as follows:
\begin{enumerate}
\item The \texttt{SceneObject} representing the entire scene is asked to return the colour of the reverse ray.
% the colour an observer positioned at the light ray's origin would see in the direction of the light ray.

\item The \texttt{SceneObject} finds the \texttt{SceneObjectPrimitive} the light ray intersects (if any), and asks its \texttt{SurfaceProperty} to return the colour of the reverse ray.

\item The \texttt{SurfaceProperty} either changes the light-ray direction and starts tracing again, or it determines the colour of the surface under illumination by the \texttt{LightSource}.
If the \texttt{SurfaceProperty} is a \texttt{SurfacePropertyContainer}, then the colours due to all the surface properties are summed (by adding up the individual RGB components).

\item The \texttt{LightSource} returns a colour according to its shading model.
If the \texttt{LightSource} is a \texttt{LightSourceContainer}, then it asks each of the \texttt{LightSource}s it contains to return a colour and then sums these colours.

\end{enumerate}
We discuss the steps in some more detail below.

In TIM, tracing an individual ray backwards is initiated by asking the \texttt{SceneObjectContainer} containing all scene objects to return the colour an observer at the ray's starting point would see in the ray's direction.
This is done by calling the \texttt{SceneObjectContainer}'s \texttt{getColour} method (which is defined in the \texttt{SceneObjectContainer}'s superclass \texttt{SceneObjectClass}).
The \texttt{SceneObjectContainer} then establishes which one (if any) of the scene objects it contains the ray would intersect with first.
If the ray intersects none of the scene objects, then the \texttt{getColour} method returns the colour black.
If the ray intersects one of the scene objects, then the method establishes which \texttt{SceneObjectPrimitive} was hit and calls this \texttt{SceneObjectPrimitive}'s \texttt{SurfaceProperty} to establish the colour.

Each \texttt{SurfaceProperty} implements a \texttt{getColour} method, which returns the colour an observer at the ray's starting point would see in the ray's direction.
In the simplest case, implemented in the class \texttt{SurfaceColourLightSourceIndependent} (surface type ``Coloured (glowing)''), the colour stored within the specific instance of the class is returned, irrespective of the illumination and the rest of the scene.
Illumination-dependent colour is handled by the \texttt{SurfaceColour} class, which calls the light source's \texttt{getColour} method to establish the appearance of the surface under illumination by the light source.
There are also surface properties, for example \texttt{Reflective}, whose \texttt{getColour} method returns the colour resulting from tracing through the scene a new ray that starts from the intersection point and travels in a new direction (given, in the case of the \texttt{Reflective} class, by the law of reflection).
Finally, it is also possible to have surface properties whose \texttt{getColour} method returns the colour resulting from tracing a new ray that starts at a point different from the intersection point; the \texttt{Teleporting} class is an example of such a class.
In the latter two cases of surface properties that continue tracing rays through the optical system, the colour may be slightly darkened to represent a reflection or transmission coefficient of less than one.

A light source's \texttt{getColour} method calculates the colour in which an observer would see a specific surface colour at a particular point on a surface if it was illuminated only by this light source.
TIM models two different types of light source: ambient light (class \texttt{AmbientLight}), which illuminates all objects in all directions with a given RGB colour; and \texttt{PhongLightSource}, which implements the Phong shading model \cite{Wikipedia-PhongShading}.
The latter corresponds roughly to a small, coloured light bulb:
it has a specific position, which is used to determine whether the surface is in another scene object's shadow and whether near-specular reflection occurs, which leads to highlights;
and it has a specific colour.
There is also a class (\texttt{LightSourceContainer}) that models the effect of combinations of different light sources.

It should be clear from the above discussion that backward tracing of a ray ends when a light ray has intersected a surface with a surface of class \texttt{SurfaceColourLightSourceIndependent}, or when it hits a surface of class \texttt{SurfaceColour} (in which case the light source performs the final calculation of colour).
Sometimes it can happen that rays get ``trapped'', for example between mirror surfaces such as those in canonical optical resonators \cite{Nelson-et-al-2008}.
In such cases, a reflection (or transmission) coefficient $<1$ ensures exponential fall-off of the intensity with the number of bounces, so such light rays become dark.
TIM limits the number of bounces it models, and when the limit is reached returns the colour black.
This is controlled by a variable called \texttt{traceLevel}, which gets passed between the objects.
The backwards tracing process starts with \texttt{traceLevel} taking a value of typically 100;
whenever a surface property initiates tracing of a new ray, it does so with a value of \texttt{traceLevel} that is reduced by 1.
When the value 0 is reached, the colour black is returned.

%\section{\label{parametrisation-appendix}Parametrisation of scene objects}
%
%
%\begin{figure}
%\begin{center} \includegraphics[width=\picturewidth]{sphereWithCoordinateSystems.pdf} \end{center}
%\caption{\label{parametrisation-figure}Parametrisation of a geometric shape using the example of a sphere of class \texttt{ParametrisedSphere}.
%Each point on the surface has an associated pair of coordinates, in this example the spherical coordinates $\theta$ and $\phi$ of the surface point, defined with respect to an arbitrary direction to the north pole (here $(0.408, 0.408, 0.816)$) and from the centre to the $0^\circ$ meridian (here $(0.913, -0.183, -0.365)$).
%This parametrisation of the surface is indicated by covering the surface in a chequerboard pattern with tiles of side length 1 in both $\theta$ and $\phi$ (surface property of class \texttt{SurfaceTiling}).
%The vectors associated with each point $\textbf{P}$ on the surface are defined as $\partial \textbf{P} / \partial \theta$ and $\partial \textbf{P} / \partial \phi$; each indicates the direction and magnitude of the change in the position $\mathbf{P}$ on the surface if one of the coordinates is changed.
%These vectors are shown for two points (using the \texttt{EditableObjectCoordinateSystem} class), namely for $(\theta, \phi) = (2.5, 1.5)$ and for $(\theta, \phi) = (2,0)$.
%In both cases, the normalised surface normal is also shown.
%The sphere has radius 1 and is centred at $(0, 0, 10)$.}
%\end{figure}

\section{\label{structure-appendix}Source-code structure}

\noindent
TIM is divided into a hierarchical package structure.
We describe here the main branch of this structure, namely the \texttt{optics} package and the packages it contains.
There are three additional packages:
\begin{enumerate}
\item \texttt{JAMA} is a matrix package in the public domain \cite{JAMA}.
TIM uses it; for convenience, the unmodified third-party source code is distributed with TIM's source code.
\item \texttt{math} is a package that defines additional mathematical functionality as required, including classes dealing with complex numbers, 2D vectors, and 3D vectors.
\item \texttt{test} contains only the class \texttt{Test}, which can be executed as a Java application for the purposes of testing any parts of the code.
\end{enumerate}

\subsection{The \texttt{optics} package}

\noindent
The \texttt{optics} package collects together optics-related code.
At the top level, it contains the \texttt{Constants} class, a collection of optics-related constants such as a few common refractive indices;
and \texttt{DoubleColour}, which is used internally to represent colours.

The only sub-package within the \texttt{optics} package that is distributed with TIM's source code is \texttt{optics.raytrace}.
It contains both the ray tracer code as well as associated mathematical, graphical and user interface code, organised in the form of a number of sub-packages.
At the top level, it contains \texttt{NonInteractiveTIM}, a template for a class that can be run as a Java application that uses TIM (see section \ref{non-interactive-TIM-appendix});
\texttt{TIMApplet}, the applet class that is called when the interactive version of TIM is run as an applet;
\texttt{TIMJavaApplication}, which allows the interactive version of TIM to be run as a Java application (with slightly increased functionality, specifically the ability to save images as \texttt{.bmp} files);
and \texttt{TIMInteractiveBits}, a class that defines the interactive parts of TIM, which are called by both the \texttt{TIMApplet} and \texttt{TIMJavaApplication} classes.

The package \texttt{optics.raytrace.core} contains a number of the core ray-tracing classes and interfaces.
A number of these, and their interactions, are discussed in \ref{ray-tracing-appendix}.
The ray-tracing core classes and interfaces include those defining the structure of cameras (\texttt{Camera}), light sources (\texttt{LightSource}; see \texttt{optics.raytrace.lights} for implementations), scene objects (\texttt{SceneObject}; implementations are in package \texttt{optics.raytrace.sceneObjects}), surface properties (\texttt{SurfaceProperty}; implementations in \texttt{optics.raytrace.surfaces}), rays (\texttt{Ray} and \texttt{RayWithTrajectory}), and intersections between rays and objects (\texttt{RaySceneObjectIntersection}).
The \texttt{Studio} class defines a collection of everything required to calculate a photo, namely information about the scene (in a \texttt{SceneObject} object), lighting (a \texttt{LightSource} object), and camera (in the form of a \texttt{Camera} object; implementations of cameras are in the package \texttt{optics.raytrace.cameras}).
A number of interfaces outline mechanisms for the parametrisation of object surfaces:
\begin{itemize}
\item \texttt{ParametrisedObject} defines coordinates assigned to each point on a surface;
\item \texttt{One2OneParametrisedObject} extends \texttt{ParametrisedObject} by asking for the reverse of the coordinate assignment, i.e.\ for a method that returns the point on a surface associated with a set of coordinate values.
\end{itemize}
Other classes included in the package are
\texttt{CameraClass}, which implements a number of methods common to currently all cameras, and which is a superclass to currently all cameras;
\texttt{SceneObjectClass}, which similarly implements a number of methods common to most scene objects, and is a superclass of many scene-object classes;
\texttt{SceneObjectPrimitive}, which describes simple geometric objects, such as spheres and planes;
\texttt{CCD} and \texttt{CentredCCD}, which represent the light-detecting element in cameras;
\texttt{Transformation}, which defines the structure of geometrical transformations (such as translation or rotation) which can be applied to scene objects (for implementations see package \texttt{optics.raytrace.sceneObjects.transformations});
and \texttt{DoubleColourCamera} and \texttt{DoubleColourCCD}, which define the structure of higher-quality cameras and their light-detecting elements.

The useful \texttt{optics.raytrace.demo} package contains a number of classes which can be run as Java applications and which demonstrate the use and effect of different features, for example \texttt{LightsDemo}, which was used to create Fig.\ \ref{lights-demo-figure}.
It is worth studying the examples contained in this package to understand how to access TIM's functionality.

The package \texttt{optics.raytrace.exceptions} defines classes that signal exceptional circumstances that occured during rendering, for example a ray becoming evanescent (\texttt{EvanescentException}).

\subsection{\texttt{optics.raytrace.cameras}}

\noindent
The \texttt{optics.raytrace.cameras} package contains implementations of the \texttt{Camera} interface.
These handle the mechanisms of generating the rays that are then traced through the scene, and of turning creating corresponding images.
Implemented camera classes include \texttt{PinholeCamera}, which is the simplest type of camera that takes pictures in which everything is in focus;
\texttt{ApertureCamera}, a camera with a circular aperture that can focus on any transverse plane;
\texttt{AnyFocusSurfaceCamera}, a camera with a circular aperture that can focus very general surfaces (section \ref{arbitrary-focus-surface-camera});
\texttt{OrthographicCamera}, which produces orthographic projections into a plane;
\texttt{AnaglyphCamera}, which can produce either red/blue or colour anaglyph images that can be viewed with standard red/cyan anaglyph glasses (section \ref{anaglyph-section});
and \texttt{AutostereogramCamera}, which can create random-dot autostereograms of the scene (section \ref{autostereogram-section}).

\subsection{\texttt{optics.raytrace.lights}}

\begin{figure}
\begin{center} \includegraphics[width=\picturewidth]{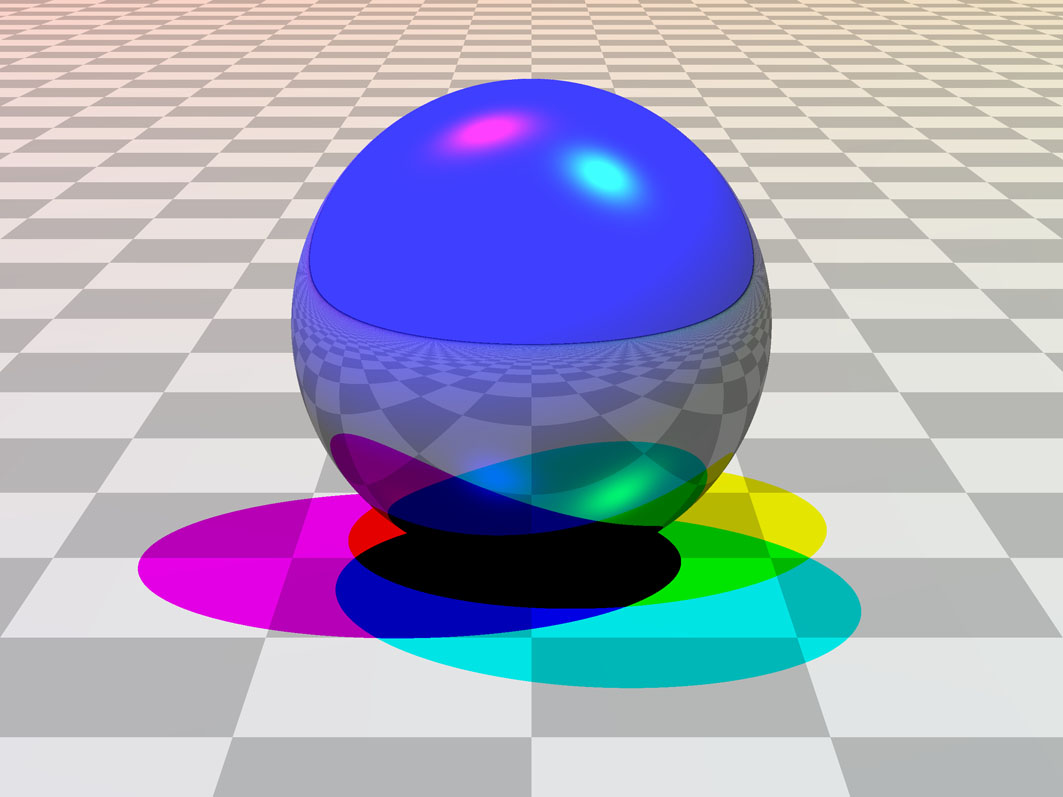} \end{center}
\caption{\label{lights-demo-figure}Effect of combination of different light sources.
Here, a shiny blue sphere is illuminated by a combination of three Phong light sources, one red, one green, one blue, placed in different directions high above the sphere.
% The ambient light ensures that no parts of the scene are completely dark.
The Phong light sources produce differently-coloured highlights at the top of the sphere and differently-coloured shadows (the colours are due to subtractive colour mixing) on the floor.
Where the shadows from all Phong light sources overlap, the scene is completely black in the absence of an ambient light source.
The image was rendered using the \texttt{LightsDemo} class in the \texttt{optics.raytrace.demo} package (\ref{structure-appendix}).}
\end{figure}

\noindent
The \texttt{optics.raytrace.lights} package includes implementations of light sources.
These include
\texttt{AmbientLight}, which represents a coloured ambient light;
\texttt{PhongLightSource}, which realizes the Phong shading model \cite{Wikipedia-PhongShading}, which is roughly equivalent to a slightly fuzzy point light source;
and \texttt{LightSourceContainer}, which allows light sources to be combined.
Fig.\ \ref{lights-demo-figure} demonstrates effects due to different light sources.

\subsection{\texttt{optics.raytrace.sceneObjects}}

\noindent
The package \texttt{optics.raytrace.sceneObjects} contains implementations of scene objects.
It contains classes describing different types of scene objects:
\begin{description}
\item [Simple geometrical shapes.]
Classes that describe simple geometrical shapes are implementations of the \texttt{SceneObjectPrimitive} class (\ref{primitive-appendix}).
Examples include spheres (\texttt{Sphere}), planes (\texttt{Plane}), parallelograms (\texttt{CentredParallelogram}), discs (\texttt{Disc}), and cylinders (\texttt{Cylinder}).

\item [Combinations of other scene objects.]
A number of classes describe compound objects (\ref{scene-object-combinations-appendix}).
Examples include \texttt{Arrow}, \texttt{Cylinder}, and \texttt{Eye}.

\item [Shapes with parametrised surfaces.]
Parametrisation, described by implementations of the interfaces \texttt{ParametrisedObject} and \texttt{One2OneParametrisedObject}, assigns coordinates to points on the surface of a geometrical shape (\ref{parametrisation-appendix}).
Examples of classes that define parametrised geometrical shapes include \texttt{ParametrisedSphere}, \texttt{ParametrisedPlane}, and \texttt{ParametrisedCentredParallelogram}.
%, \texttt{ParametrisedDisc}, and \texttt{ParametrisedCylinder}.
There are also classes that allow the range of the coordinates that describes the surface of the geometrical shape to be varied.
For example, in the \texttt{ScaledParametrisedSphere} class, the range of the polar angle $\theta$ can be set to range from an arbitrary value $\theta_\mathrm{min}$ to another arbitrary value $\theta_\mathrm{max}$.
Other examples include \texttt{ScaledParametrisedDisc} and \texttt{ScaledParametrisedCentredParallelogram}.
\end{description}

The \texttt{optics.raytrace.sceneObjects.solidGeometry} package is a collection of classes useful for combining scene objects.
In the simplest case, scene objects are grouped in a hierarchical way (\texttt{SceneObjectContainer}).
More elaborate combinations of scene objects include intersections (\texttt{SceneObjectIntersection}), unions (\texttt{SceneObjectUnion}), and inverse (\texttt{SceneObjectInverse}).

One of the capabilities required of any scene object is the ability to create a transformed copy of itself.
The structure of the transformation is defined by the \texttt{Transformation} class (in \texttt{optics.raytrace.core}).
The package \texttt{optics.raytrace.sceneObjects.transformations} contains classes that describe specific types of transformation, i.e.\ subclasses of \texttt{Transformation}.
Examples include \texttt{Translation}, \texttt{RotationAroundXAxis}, \texttt{RotationAroundYAxis},  \texttt{RotationAroundZAxis}, and the more general \texttt{LinearTransformation}.

\subsection{\texttt{optics.raytrace.surfaces}}

\noindent
The \texttt{optics.raytrace.surfaces} package contains implementations of the \texttt{SurfaceProperty} interface (in \texttt{optics.raytrace.core}).
These include
\begin{itemize}
\item the classes representing coloured surfaces, \texttt{SurfaceColourLightSourceIndependent} and \texttt{SurfaceColour};
\item a class representing a transparent surface (\texttt{Transparent});
\item the \texttt{Reflective} class which represents specularly reflective surfaces;
\item the \texttt{Refractive} class which represents refraction at the interface between media with different refractive indices according to Snell's law;
\item a class that facilitates the implementation of classes that represent surfaces that change direction according to generalised laws of refraction (\texttt{Metarefractive});
\item a number of subclasses of \texttt{Metarefractive} representing surfaces that invert of one of the ray-direction components tangential to the surface \cite{Hamilton-Courtial-2008a} (\texttt{RayFlipping}), rotate the ray direction around the local surface normal \cite{Hamilton-et-al-2009} (\texttt{RayRotating}), and refract like --- formally \cite{Sundar-et-al-2009} --- the interface between media with a \emph{complex} refractive-index ratio would (\texttt{RefractiveComplex});
\item classes representing surfaces that have combinations of other surface properties (\texttt{SurfacePropertyContainer} and \texttt{SurfacePropertyContainerWeighted});
\item a \texttt{SemiTransparent} class, which combines an arbitrary surface property with the \texttt{Transparent} surface property;
\item a class representing a surface whose inside has different properties from its outside (\texttt{TwoSidedSurface});
\item classes representing surfaces with completely different surface properties at different points on the surface (\texttt{SurfaceTiling}, \texttt{EitherOrSurface}, \texttt{PictureSurface}, \texttt{PictureSurfaceDiffuse}, \texttt{PictureSurfaceSpecular};
\item a class representing the hologram of a thin lens (\texttt{ThinLensHologram}), which changes light-ray direction dependent on the point where it is being intersected;
\item and a class representing a surface a light ray enters and then continues, from a corresponding position and with a corresponding direction, from the surface of a different scene object (\texttt{Teleporting} --- see section \ref{teleporting-section}; we use this to model geometric optical transformations \cite{Berkhout-et-al-2010}).
\end{itemize}
\ref{new-surface-property-appendix} discusses how to add new surface properties to TIM.

The \texttt{optics.raytrace.surfaces.metarefraction} package contains description of the direction change performed by surfaces of class \texttt{Metarefractive} (see \texttt{optics.raytrace.surfaces}).
The format of these descriptions is defined by the abstract \texttt{Metarefraction} class.
The \texttt{ComplexMetarefraction} class is a subclass of \texttt{Metarefraction}, again abstract, which formulates the direction change in terms of the projections of the incoming and outgoing light rays into an Argand plane tangential to the surface at the intersection point and with its origin there (section \ref{surface-property-section}) \cite{Constable-et-al-2011}.
This is a generalisation of the formal description in terms of multiplication with a complex number of rotation of the light-ray direction around the local surface normal \cite{Sundar-et-al-2009}.
All other classes in this package are non-abstract subclasses of \texttt{ComplexMetarefraction}.
They include classes that describe surfaces that change light-ray direction described by complex multiplication (\texttt{ComplexMetarefractionMultiplication}), complex addition (\texttt{ComplexMetarefractionAddition}), complex conjugation (\texttt{ComplexMetarefractionCC}), and complex exponentiation (\texttt{ComplexMetarefractionExp}).

\subsection{\texttt{optics.raytrace.GUI}}

\noindent
The \texttt{optics.raytrace.GUI} package is a collection of packages that together constitute TIM's graphical user interface (GUI).
All but one of the sub-packages of \texttt{optics.raytrace.GUI} contain classes that handle user interaction in TIM's interactive version.

The exception that contains the classes that handle user interaction in the non-interactive version of TIM is the package \texttt{optics.raytrace.GUI.nonInteractive}.
It contains two classes:
\texttt{PhotoCanvas}, which creates a panel (screen element) in which \texttt{NonInteractiveTIM} and classes derived from it display the rendered image;
and \texttt{PhotoFrame}, which opens a window containing a \texttt{PhotoCanvas}.

\begin{figure}
\begin{center} \includegraphics[width=\picturewidth]{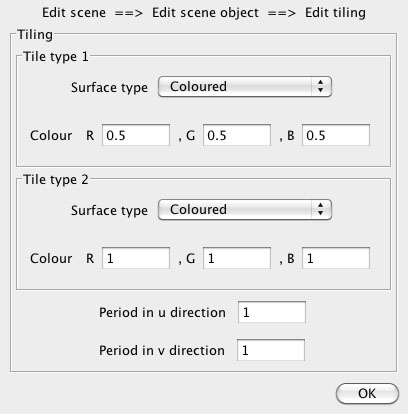} \end{center}
\caption{\label{IPanel-figure}Editing one object in a hierarchical structure using the \texttt{IPanel} class.
The top line displays a breadcrumb trail, giving an idea of the place in the hierarchy.}
\end{figure}

The package \texttt{optics.raytrace.GUI.core} contains GUI core classes.
The \texttt{RaytraceWorker} class handles rendering in a background thread, so that the GUI can continue to react to user interaction.
The class \texttt{IPanel} defines a screen element intended to allow browsing and editing hierarchical networks of objects on a relatively small screen area, so that it can be integrated into an internet page without necessarily dominating the page.
At any one time, the panel corresponding to one object in this network is displayed.
\texttt{IPanel} handles a stack of these panels, displaying only the top one, and a breadcrumb trail of the other panels in the stack (Fig.\ \ref{IPanel-figure}).
The mechanism by which this panel is supplied by the object, and how control is handed over when another object's panel gets displayed in the \texttt{IPanel}, is defined in the \texttt{IPanelComponent} interface, which all objects in TIM which are editable interactively implement.
\texttt{EditableCamera} is an interface that defines the functionality of an editable camera.

The package \texttt{optics.raytrace.GUI.lowLevel} is a collection of low-level classes related to the GUI.
The class \texttt{GUIFrame} represents the window TIM's interactive version opens when run as a Java application.
\texttt{GUIPanel} is the top-level screen element that contains the entire GUI, i.e.\ the tabs showing views of the rendered scene from different view points, any component being edited, and all buttons (see Fig.\ \ref{TIM-window-figure}).
\texttt{BufferedImageCanvas} is the panel displaying the rendered image.
\texttt{RaytracingImageCanvas} extends \texttt{BufferedImageCanvas} by adding some interactive functionality, including the ability to identify the scene object and coordinates of the point on the object the mouse pointer hovers over and clicks on, and the ability to edit that scene object by double-clicking on it.
Most classes provide panels for editing small chunks of data, including
integers (\texttt{IntPanel} and \texttt{LabelledIntPanel}, an \texttt{IntPanel} displayed next to a brief verbal description),
individual double-precision real numbers (\texttt{DoublePanel} and \texttt{LabelledDoublePanel}),
pairs of double-precision real numbers (\texttt{TwoNumbersPanel}),
complex numbers (\texttt{ComplexPanel} and \texttt{LabelledComplexPanel}),
2D vectors (\texttt{Vector2DPanel} and \texttt{LabelledVector2DPanel}),
3D vectors (\texttt{Vector3DPanel} and \texttt{LabelledVector3DPanel}),
and the limiting values of a range of real numbers (\texttt{LabelledMinMaxPanel}).
The class \texttt{SceneObjectPrimitivesComboBox} describes a panel that allows any \texttt{SceneObjectPrimitive} in the scene to be selected, which is used in the panel editing surface properties for selecting a target object for a surface of type \texttt{Teleporting}.
There are a number of classes for editing camera-specific parameters such as aperture size (\texttt{ApertureSizeComboBox} and \texttt{LabelledApertureSizeComboBox});
quality (\texttt{QualityComboBox} and \texttt{LabelledQualityComboBox}), which can be applied to blurring and to anti-aliasing;
and all parameters related to camera blur collected in one panel (\texttt{BlurPanel}).
The class \texttt{ButtonsPanel} describes a panel containing one or more buttons.
The interface \texttt{StatusIndicator} outlines the structure of an object that can display brief status messages.
Finally, \texttt{GUIBitsAndBobs} is a small collection of miscellaneous methods commonly used by the GUI.

The package \texttt{optics.raytrace.GUI.cameras} contains classes that describe various editable camera classes.
The classes \texttt{EditableAnyForucSurfaceCamera}, \texttt{EditableApertureCamera}, \texttt{EditableAnaglyphCamera}, \texttt{EditableAutostereogramCamera} and \texttt{EditableOrthographicCamera} respectively extend the \texttt{AnyFocusSurfaceCamera}, \texttt{ApertureCamera}, \texttt{AnaglyphCamera}, \texttt{AutostereogramCamera} and \texttt{OrthographicCamera} classes in this way.
The classes \texttt{EditableOrthographicCameraSide} and \texttt{EditableOrthographicCameraTop} make special cases of the \texttt{OrthographicCamera} class editable; these respectively correspond to TIM's ``Side view'' and ``Top view'' tabs.

The package \texttt{optics.raytrace.GUI.sceneObjects} contains the classes describing all editable scene objects.
Most of these are simply editable versions of classes in \texttt{optics.raytrace.sceneObjects}, including \texttt{EditableArrow}, \texttt{EditableParametrisedCone}, \texttt{EditableParametrisedCylinder}, \texttt{EditableParametrisedPlane}, \texttt{EditableRayTrajectory}, \texttt{EditableRayTrajectoryCone}, \texttt{EditableScaledParametrisedDisc}, \texttt{EditableScaledParametrisedCentredParallelogram}, and \texttt{EditableScaledParametrisedSphere}.
The class \texttt{EditableSceneObjectCollection} allows editing of groups of scene objects, which can be combined a number of ways respectively handled by the \texttt{SceneObjectContainer}, \texttt{SceneObjectIntersection}, and \texttt{SceneObjectUnion} classes in the \texttt{optics.raytrace.sceneObjects.solidGeometry} package.
A few of the classes defined in \texttt{optics.raytrace.GUI.sceneObjects} exist only in editable form.
Examples include \texttt{EditableCylinderFrame}, \texttt{EditableCylinderLattice}, \texttt{EditableLens}, \texttt{EditableObjectCoordinateSystem}, \texttt{EditablePolarToCartesianConverter}, and \texttt{EditableTelescope}.

The \texttt{optics.raytrace.GUI.surfaces} package contains classes that enable selecting a class of surface property and editing class-specific parameters.
The class that provides a panel for doing all of this is \texttt{SurfacePropertyPanel} (see Fig.\ \ref{SurfacePropertyPanel-figure}, appendix \ref{interactive-surface-property-appendix}).
The remaining classes in this package allow editing of class-specific parameters.
\texttt{EditableSurfaceTiling} and \texttt{EditableTwoSidedSurface} are editable subclasses of \texttt{SurfaceTiling} and \texttt{TwoSidedSurface}, respectively.
\texttt{TeleportingTargetsComboBox} is a subclass of \texttt{SceneObjectPrimitivesComboBox} (in the \texttt{optics.raytrace.GUI.lowLevel} package) that allows a suitable scene object contained in the scene to be selected.
In its labelled form (\texttt{LabelledTeleportingTargetsComboBox}) this is used in the \texttt{SurfacePropertyPanel} class to select a target object for the \texttt{Teleporting} surface property.

Last, and least, the package \texttt{optics.raytrace.GUI.sceneObjects.transformations} contains only the class \texttt{EditableLinearTransformation}, which will eventually be able to edit linear scene-object transformations (see package \texttt{optics.raytrace.sceneObjects.transformations}) and become part of TIM's interactive version.

\section{\label{non-interactive-TIM-appendix}The default non-interactive TIM}

\noindent
TIM's source code comes with a class called \texttt{NonInteractiveTIM}, which can be compiled and run as a Java application\footnote{In Eclipse \cite{Eclipse}, simply bring up the source-code file \texttt{optics.raytrace.RayTraceJavaApplication.java} and select \texttt{Run > Run}.} that defines a studio (scene, camera and lights);
renders the scene under the conditions defined in the studio;
and displays the rendered image on the screen and saves  it as a \texttt{.BMP} file.
This class can serve as an example and template for using TIM's source code.

The class \texttt{NonInteractiveTIM} provides three methods:
\begin{enumerate}
\item \texttt{getFilename} returns the filename under which the rendered image is saved;
\item \texttt{createStudio} defines and returns the studio, i.e.\ a scene, a camera, and lights;
\item \texttt{main} calls the \texttt{createStudio} method, renders the image, displays it on the screen, and saves it as a \texttt{.BMP} file;  automatically gets called when the \texttt{NonInteractiveTIM} class is run as a Java application.
\end{enumerate}
Modifying these methods\footnote{Note that overriding these methods has no effect as all are declared \texttt{static}.} % The reason is that the \texttt{main} method has to be declared \texttt{static}; the other methods are called from within it, without instantiating any \texttt{NonInteractiveTIM} object.
allows TIM to perform tasks that cannot currently be achieved in the interactive version; below are a few examples.

First we discuss modifying the \texttt{createStudio} method, which changes one or more of scene, camera and lights.
This allows the programmer to do a number of things that are not currently possible in the interactive version, including
\begin{itemize}
\item setting parameter values by typing in formulas;
\item using Java's built-in flow control (such as \texttt{for} or \texttt{while} loops) and formulas to create a number of objects systematically;
\item accessing classes of scene object not currently supported in the interactive version (e.g.\ \texttt{MaskedObject});
% \item accessing scene objects that are parametrised differently (e.g.\ \texttt{ParametrisedSphere2}, which ...);
\item transforming objects, e.g.\ rotate, move, or scale them (these transformations are defined in the \texttt{optics.raytrace.sceneObjects.transformations} package);
\item accessing additional classes of surface properties (an example is \texttt{PictureSurface}, which maps a  picture loaded from a file onto a surface; other examples include several types of \texttt{MetarefractiveSurface});
\item giving surfaces \emph{combinations} of surface properties (which can be achieved by using the \texttt{SurfacePropertyContainer} or \texttt{SurfacePropertyContainerWeighted} classes);
\item changing the lights.
\end{itemize}

Changing the behaviour of the resulting Java application altogether can be achieved by altering the \texttt{main} class.
Simple examples include stopping the Java application from saving the image (by removing the relevant line in the code, or simply by commenting it out).
By running a loop in which parameters change values, for example the position of a scene object or the camera, and by saving the rendered images with a suitable filename (in the simplest case ending in a number), the saved images can later be combined into a movie by other software.

\section{\label{new-scene-object-appendix}Adding a new scene-object class}

\noindent
Sometimes it is necessary to add a new class of scene object to TIM.
We can distinguish the following cases, which we treat in more detail in the following sections:
\begin{enumerate}
\item It is desirable to add a class representing a geometrical shape not yet represented in TIM, for example an ellipsoid or a torus.
\item It is desirable to define a class representing combinations of geometrical shapes already represented in TIM.
One reason for doing so could be to automate the placing and adding to the scene of the constituent scene objects, for example the cone and the cylinder that form an arrow represented by the \texttt{Arrow} class.
% cylinders in an array of cylinders represented by the \texttt{EditableCylinderLattice} class.
The individual scene objects can also automatically be combined using the solid-geometry classes defined in the \texttt{optics.raytrace.sceneObjects.solidGeometry} package; this is how a lens is created by the \texttt{EditableLens} class.
\item It is desirable to (re)parametrise the surface of an existing geometrical shape.
\item It is desirable to add a scene-object class that is represented in TIM for non-interactive use to the interactive version.
\end{enumerate}

\subsection{\label{primitive-appendix}Adding a class representing a geometrical shape}

\noindent
A geometrical shape is represented by a (non-abstract) subclass of the (abstract) \texttt{SceneObjectPrimitive} class.
This is itself a subclass of \texttt{SceneObjectClass}, an abstract class that implements some common methods required by the \texttt{SceneObject} interface such as keeping a copy of the studio, the parent object, and the description, and providing implementations of methods such as \texttt{getColour} that reduce the task to other methods that remain to be implemented, such as finding the intersection between a ray and the object.
It is instructive to study the implementation of such a class, for example \texttt{Sphere}.

Any new subclass of the \texttt{SceneObjectPrimitive} class needs to implement methods for dealing with the geometry of ray tracing, namely finding the intersection between a ray and the shape (\texttt{getClosestRayIntersection}); calculating the normalised surface normal at any possible intersection point (\texttt{getNormalisedSurfaceNormal}); and determining whether a position is inside the object or not (\texttt{insideObject}).
It also needs to implement methods for an instance of the class to make an identical copy of itself (\texttt{clone}), or a copy that is geometrically transformed, for example shifted, rotated, or scaled (\texttt{transform}).

Note that the surface normal points in the direction of the shape's outside.
In many cases, it is obvious which side of the surface is on the inside and which one is on the outside, for example in the case of a sphere.
However, in other cases, for example in the case of a plane, it is not at all obvious.
In such cases, the direction of the surface normal \emph{defines} an inside and an outside.
It is important to be able to distinguish inside from outside as many types of surface property, for example refraction, distinguish between rays that arrive from the inside from those arriving from the outside.

\subsection{\label{scene-object-combinations-appendix}Defining a new scene-object class in terms of existing scene objects}

\noindent
Sometimes it is desirable to define a new class of scene objects that consists of a number of scene objects of existing classes.
Examples include the \texttt{Arrow} class, which represents a combination of a cone (the arrow's tip) and a cylinder (the shaft), and the \texttt{EditableLens} class, which represents a convex-convex lens.

The easiest way to implement such a class is to extend the class that represents the appropriate combination of scene objects.
For example, an arrow, which is simply a collection of a cone and a cylinder, can be realised by extending the class representing simple scene-object collections, \texttt{SceneObjectContainer};
a convex-convex lens, which is the intersection of two spheres, can be realised by extending the \texttt{SceneObjectIntersection} class.
All the class then needs to do (usually in the constructor) is to add the appropriate scene objects to the array of objects in the class, using the \texttt{addSceneObject} method.

\subsection{\label{parametrisation-appendix}(Re)parametrising the surface of an existing shape}

\noindent
A number of surface properties require a surface that has a two-dimensional coordinate system associated with it (section \ref{teleporting-section}, especially Fig.\ \ref{parametrisation-figure}).
An example of such a surface property is \texttt{SurfaceTiling}, which covers the surface in a chequerboard pattern in the surface's coordinate system.
A surface that can calculate a pair of coordinates for any point on the surface is represented by the \texttt{ParametrisedObject} interface; it needs to implement methods for returning coordinates of an arbitrary point on the surface (\texttt{getSurfaceCoordinates}) and for returning the corresponding coordinate names, for example \texttt{theta} and \texttt{phi} in the case of a sphere's polar coordinate system (\texttt{getSurfaceCoordinateNames}).
A surface that can also identify a point on the surface for arbitrary coordinates is represented by \texttt{One2OneParametrisedObject} interface.
Such a surface needs to implement the methods required by the \texttt{ParametrisedObject} interface, and additionally the \texttt{getPointForSurfaceCoordinates} method.
If the mapping between surface points and coordinates is not one-to-one then the behaviour of this method is undefined.

Note that there are scene objects, most notably compound scene objects, which can implement the \texttt{ParametrisedObject} interface so that patterned surfaces can be applied to each constituent scene object, provided it implements \texttt{ParametrisedObject}, but which should not implement the \texttt{One2OneParametrisedObject} interface even if all constituent scene objects do as there might be points on different constituent objects that correspond to the same combination of coordinate values, and so the mapping between surface points and coordinates is not one-to-one.

%\begin{figure}
%\begin{center} \includegraphics[width=\picturewidth]{sphereWithCoordinateSystems.jpg} \end{center}
%\caption{\label{parametrisation-figure}Parametrisation of a geometric shape using the example of a sphere of class \texttt{ParametrisedSphere}.
%Each point on the surface has an associated pair of coordinates, in this example the spherical coordinates $\theta$ and $\phi$ of the surface point, defined with respect to an arbitrary direction to the north pole (here $(0.408, 0.408, 0.816)$) and from the centre to the $0^\circ$ meridian (here $(0.913, -0.183, -0.365)$).
%This parametrisation of the surface is indicated by covering the surface in a chequerboard pattern with tiles of side length 1 in both $\theta$ and $\phi$ (surface property of class \texttt{SurfaceTiling}).
%The vectors associated with each point $\textbf{P}$ on the surface are defined as $\partial \textbf{P} / \partial \theta$ and $\partial \textbf{P} / \partial \phi$; each indicates the direction and magnitude of the change in the position $\mathbf{P}$ on the surface if one of the coordinates is changed.
%These vectors are shown for two points (using the \texttt{EditableObjectCoordinateSystem} class), namely for $(\theta, \phi) = (2.5, 1.5)$ and for $(\theta, \phi) = (2,0)$.
%In both cases, the normalised surface normal is also shown.
%The sphere has radius 1 and is centred at $(0, 0, 10)$.}
%\end{figure}

A scene object that implements the \texttt{ParametrisedObject} interface also defines directions on the surface, through the \texttt{getSurfaceCoordinateAxes} call-back method.
This allows the implementation of anisotropic surface properties, which require a preferred direction to be defined.
Fig.\ \ref{parametrisation-figure} shows these vectors for two points on the surface of an object of class \texttt{ParametrisedSphere}.
Their primary purpose is to define directions on the surface, which is why they many methods that use these vectors normalise them.
An example is the surface-property class \texttt{Metarefractive}, which allows surfaces to deflect light rays in very general ways.
(This can be seen as a generalisation of refraction at the interface between media with different refractive indices, or ``metarefraction'' \cite{Hamilton-Courtial-2009}.)
As many light-ray-direction changes are internally handled through the \texttt{Metarefractive} class (see \ref{new-metarefractive-surface-property-appendix}), it is often important that surfaces implement the \texttt{ParametrisedObject} interface.
Those light-ray-direction changes include Snell's-law refraction (surface property \texttt{Refractive});
ray flipping, which changes the sign of one light-ray-direction component \cite{Hamilton-Courtial-2008a} (surface property \texttt{RayFlipping});
and ray rotation (Fig.\ \ref{TIM-window-figure}), which rotates the light-ray direction by an arbitrary (but fixed) angle around the local surface normal \cite{Hamilton-et-al-2009} (surface property \texttt{RayRotating}).

% One surface property for which not only the direction of the vectors associated with a surface point is important but also their length is \texttt{Teleporting}.

Creating a class that parametrises the geometrical shape described in an existing class can be achieved by creating a subclass of the existing class.
This new subclass needs to implement the methods of the relevant interfaces, \texttt{ParametrisedObject} or \texttt{One2OneParametrisedObject}.
If the existing class is already parametrised and the new subclass overrides its methods related to parametrisation, then the new class re-parametrises the geometrical shape described in the existing class.

\subsection{Adding an existing scene-object class to interactive TIM}

\noindent
All scene-object classes available in the interactive version of TIM need to be fully parametrised (see previous section), i.e.\ they need to implement the  \texttt{One2OneParametrisedObject} interface.

For a class to be editable through the mechanism built into the interactive version of TIM, it needs to implement the \texttt{IPanelComponent} interface.
To add an existing scene-object class to the interactive version of TIM, it is easiest to create a subclass that implements the following methods required by \texttt{IPanelComponent}:
\begin{enumerate}
\item \texttt{createEditPanel} prompts the creation of the edit panel (of class \texttt{JPanel}), an area of screen that allows interactive editing of any of the scene object's parameters;
\item \texttt{discardEditPanel} signals to the \texttt{IPanelComponent} that the edit panel is no longer required, and any resources associated with it can be freed up;
\item \texttt{getEditPanel} returns the edit panel;
\item \texttt{setValuesInEditPanel} sets all the sub-panels in the edit panel to reflect the current values of the scene object's parameters;
\item \texttt{acceptValuesInEditPanel} sets the scene object's parameters to the edited values in the edit panel;
\item \texttt{backToFront} gets invoked after editing of a different \texttt{IPanelComponent}'s edit panel has been completed (for example, the surface property of the current scene object could have been edited), and this one's edit panel is being edited again.
\end{enumerate}

The new class can now be part of the scene, and it can be edited, but it is so far not possible to create a new instance (unless another one is being duplicated).
Creating a new instance of a scene object in interactive TIM happens by adding a new scene object to a collection of scene objects.
This collection can be ``The scene'', which is the top-level collection of scene objects, but it can also be a collection that is part of ``The scene'' (or of other collections in the hierarchy).
Editing a collection is handled by the \texttt{EditableSceneObjectCollection} class.
We will discuss the required steps using the example of the \texttt{EditableTelescope} class.
\begin{enumerate}

\item Add a string that describes an object of this class (e.g.\ \texttt{String OBJECT\_TELESCOPE = "Telescope";}).

\item Add this string to the array of strings that are to be the menu items in the combo box responsible for initiating the creation of a new instance of a scene object (e.g.\ \texttt{String[] componentStrings = $\{$ [...], OBJECT\_TELESCOPE, [...]$\}$;})

\item In the \texttt{actionPerformed} method of the internal class \texttt{SceneObjectControlPanel}, which gets called whenever the user interacts with the combo box that initiates the creation of a new scene object, add a case that gets invoked when the user has selected the new object type in the new-object combo box.
In the case of our example, it has the following form:
\begin{verbatim}
else if(newElementName.equals(
    OBJECT_TELESCOPE)
  )
  iPanelComponent = new EditableTelescope(
      "Telescope",  // description
      // default centre
      new Vector3D(0, 0, 10),
      // default ocular normal
      new Vector3D(0, 0, -1),
      1,  // default magnification
      1,  // default radius of aperture
      // parent in hierarchy
      EditableSceneObjectCollection.this,  
      // scene, lights and camera
      getStudio()
    );
\end{verbatim}
\end{enumerate}

\section{\label{add-surface-property-appendix}Adding a surface-property class}

\noindent
One of the main reasons why we wrote TIM was to be able to fully control the effect of surfaces on light rays, which has proved tremendously useful not only for our research on METATOYs but also for our work on optical orbital angular momentum.
This section outlines how new surface properties can be added to TIM.

\subsection{\label{new-surface-property-appendix}Adding a general surface property}

\noindent
TIM establishes the effect of a surface on any specific light ray by asking the \texttt{getColour} method of the \texttt{SurfaceProperty} object representing the surface to return the colour of the reverse ray, i.e.\ the light ray travelling in the opposite direction.
The role of surfaces in establishing this colour is outlined in more detail in \ref{ray-tracing-appendix}.

A new surface property can be created by implementing the \texttt{SurfaceProperty} interface directly.
The class representing the new surface property must implement the \texttt{getColour} method.
Precisely how it calculates the reverse ray's colour varies greatly between different surface-property classes.
The \texttt{getColour} method has access to all the information being passed to it as arguments:
\begin{itemize}
\item the \texttt{Ray} object describing the incident light ray contains its direction and starting position;
\item the \texttt{RaySceneObjectIntersection} object describing the intersection between the surface and the light ray contains the position of the intersection point and the primitive scene object being intersected;
\item the entire scene information is passed in the form of a \texttt{SceneObject} object;
\item information about lights is passed in the form of a \texttt{LightSource} object.
\end{itemize}
Additionally, the trace level is passed as an argument.
The class might also store additional information and make it available to be used within \texttt{getColour}.

It might be instructive to study the code of a few classes that implement the \texttt{SurfaceProperty} interface directly.
The following two classes are perhaps particularly instructive.
\begin{itemize}
\item The class \texttt{SurfaceTiling} is an example of a spatially varying surface property that calculates the local coordinates in the intersected primitive scene object's coordinate system.
This is done by the code fragment
\begin{verbatim}
(
  (ParametrisedObject)(i.o)
).getSurfaceCoordinates(i.p),
\end{verbatim}
where \texttt{i} is the \texttt{RaySceneObjectIntersection} object describing the intersection point; \texttt{i.o} is the primitive scene object being intersected (which has to implement the \texttt{ParametrisedObject} interface here; if it does not, an exception is thrown);
and \texttt{i.p} is a 3D vector describing the position of the intersection point.

\item The class \texttt{Reflective} is an example of a surface property that requires further ray tracing.
Its \texttt{getColour} method illustrates, amongst other things, checking that the trace level is greater than zero and continuing backwards ray tracing with a new direction.
The latter is achieved by the following code segment:
\begin{verbatim}
return scene.getColourAvoidingOrigin(
  ray.getBranchRay(i.p, newRayDirection),
  // the primitive scene object being
  // intersected
  i.o,
  l,    // the light source(s)
  scene,    // the entire scene
  traceLevel-1
).multiply(reflectionCoefficient);
\end{verbatim}
Note that creating the continuation of the ray using the original ray's \texttt{getBranchRay} method ensures the ray trajectory is recorded correctly (see section \ref{light-ray-visualisation-section});
that the branch ray is being launched with a trace level reduced by 1;
and that the intensity of the branch ray is multiplied by a reflection coefficient stored in the \texttt{Reflective} object.
\end{itemize}

\subsection{\label{new-metarefractive-surface-property-appendix}Adding a surface property that affects only light-ray direction}

\noindent
As surface properties that change light-ray direction in different ways are of particular interest to us for the purposes of our METATOYs research, we have put in place a surface-property class that aims to facilitate the creation of new light-ray-direction-changing surface properties.
The direction change itself is described by a subclass of the abstract \texttt{Metarefraction} class, which must provide methods that calculate the new light-ray direction from the old light-ray direction for the two cases of light travelling inwards, i.e.\ arriving from the surface's outside, or outwards, i.e.\ arriving from the inside.
(Inside and outside are defined by the direction of the normalised surface normal, which can be obtained with the \texttt{SceneObjectPrimitive} class's \texttt{getNormalisedSurfaceNormal} method, and which is defined to point outwards.)
These methods are the call-back methods \texttt{refractInwards} and \texttt{refractOutwards}.
The surface-property class that represents this light-ray-direction change is of class \texttt{Metarefractive}.
The description of the direction change (i.e.\ the implementation of the \texttt{Metarefraction} class) is passed as an argument to the constructor of the \texttt{Metarefractive} surface property;
in other words, an object of class \texttt{Metarefractive} is always created around a specific light-ray-direction change. 

In TIM, most light-ray-direction changes at surfaces are described in terms of their projection into an Argand plane tangential to the surface at the intersection point and with its origin placed there (section \ref{surface-property-section}) \cite{Constable-et-al-2011}.
A complex number that corresponds to the projection of the incoming light ray is mapped to another complex number, which corresponds to the projection of the outgoing light ray.
The component in the direction of the surface normal is calculated so that the length of the direction vector remains unchanged.
The mapping of the light-ray projection then defines the direction change;
for example, refraction according to Snell's law is described by multiplication with a real number, and rotation through an angle $\alpha$ around the surface normal is described by multiplication with a complex number of the form $\exp(\rmi \alpha)$ \cite{Sundar-et-al-2009}.
Direction changes can be described in this way by the \texttt{ComplexMetarefraction} class, which is a subclass of \texttt{Metarefraction}.

Examples of surface properties which have been implemented by extending the \texttt{Metarefractive} surface-property class, and which use extensions of the \texttt{ComplexMetarefraction} class to describe the light-ray-direction change they represent, include \texttt{RayFlipping}, which represents surfaces that change the sign of one of the ray-direction components tangential to the surface at the intersection point \cite{Hamilton-Courtial-2008a};
\texttt{RayRotating}, which represents surfaces that rotate the light-ray direction by an arbitrary, but fixed, angle around the local surface normal \cite{Hamilton-et-al-2009};
\texttt{Refractive}, which represents standard Snell's-law refraction at the interface between optical media with different refractive indices;
and \texttt{RefractiveComplex}, which represents a combination of Snell's-law refraction and rotation around the local surface normal, which can be described formally as refraction at the interface between optical media with different \emph{complex} refractive indices~\cite{Sundar-et-al-2009}.

\subsection{\label{interactive-surface-property-appendix}Adding an existing surface-property class to interactive TIM}

\begin{figure}
\begin{center} \includegraphics[width=\picturewidth]{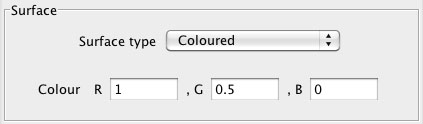} \end{center}
\caption{\label{SurfacePropertyPanel-figure}The panel for editing the surface property an orange object.
The top half allows selecting a surface-property class;
the lower half allows editing of parameters specific to the selected class, in this case the red (R), green (G), and blue (B) values of the surface colour.
The panel is created by the class \texttt{SurfacePropertyPanel}.}
\end{figure}

\noindent
Surface properties are being selected in a panel described by the \texttt{SurfacePropertyPanel} class (Fig.\ \ref{SurfacePropertyPanel-figure}).
This panel consists of a combo box that allows selection of the class of surface property, and space for editing any parameters specific to the selected surface-property class.

The following modifications of the \texttt{SurfacePropertyPanel} class add an existing surface-property class to TIM's interactive incarnation; we discuss these using the example of the surface property describing a coloured surface (\texttt{SurfaceColour} in \texttt{optics.raytrace.surfaces}).
\begin{enumerate}
\item Define a string constant, with a suitable name, that describes the surface property.
After completion of the steps below, the contents of this string will come up as an option in the combo box for selecting a surface-property class.
In our example, add the lines
\begin{verbatim}
private static final String
  SURFACE_PROPERTY_DESCRIPTION_COLOURED =
    "Coloured";
\end{verbatim}

\item If the surface-property class requires additional parameters, define a variable that can hold the panel for editing these parameters.
In our example, this panel allows separate editing of the red, green and blue (RGB) components of the surface colour, and it has a label which helpfully points out what it is the user is editing (``Colour'').
All of this can be achieved with the \texttt{LabelledDoubleColourPanel} class, using the following lines of code:
\begin{verbatim}
private LabelledDoubleColourPanel
  colourPanel;
\end{verbatim}
In the constructor of the \texttt{SurfacePropertyPanel} class, create an instance of the panel for the additional parameters and initialise it with a default value:
\begin{verbatim}
colourPanel =
  new LabelledDoubleColourPanel("Colour");
colourPanel.setDoubleColour(
    DoubleColour.WHITE
  );
\end{verbatim}

\item The \texttt{setSurfaceProperty} method contains a chain of \texttt{if} statements that distinguishes between different surface-property classes of the variable \texttt{surfaceProperty}.
Into this chain, link a case for the surface-property class to be added.
In the code block for this \texttt{if} statement, add statements that make the surface-property class that currently selected in the surface-property-class combo box;
set the panel for editing the surface-property-class-specific parameters to reflect the properties of the variable \texttt{surfaceProperty};
and ensure that the panel for editing the class-specific parameters is shown.
The following block of code does this for our example: 
\begin{verbatim}
if(surfaceProperty instanceof SurfaceColour)
{
  surfacePropertyComboBox.
    setSurfacePropertyString(
      SURFACE_PROPERTY_DESCRIPTION_COLOURED
    );
  colourPanel.setDoubleColour(
    (
      (SurfaceColour)surfaceProperty
    ).getDiffuseColour()
  );
  setOptionalParameterPanelComponent(
      colourPanel
    );
}
\end{verbatim}

\item There is a similar chain of \texttt{if} statements in the \texttt{getSurfaceProperty} method.
Into that chain, add a case that returns an instance of the new surface-property class with the parameters from the class-specific-parameters panel.
In our case, the following lines do this:
\begin{verbatim}
if(surfacePropertyString.equals(
    SURFACE_PROPERTY_DESCRIPTION_COLOURED
  ))
{
  // return a shiny version of the colour
  return new SurfaceColour(
      colourPanel.getDoubleColour(),
      // specular component; white = shiny
      DoubleColour.WHITE
  );
}
\end{verbatim}

\item The remaining changes are to the internal class \texttt{SurfacePropertyComboBox}, which describes the combo box for selecting a surface-property class.
First, the code for adding the string describing the new surface-property class to the various options available for selection by the combo box needs to be added to the constructor.
This is done by adding the string constant describing the surface property defined above to the array of strings called \texttt{surfacePropertyStrings}.
For our example,
\begin{verbatim}
surfacePropertyStrings.add(
    SURFACE_PROPERTY_DESCRIPTION_COLOURED
  );
\end{verbatim}

\item Finally, in the \texttt{actionPerformed} method, a case needs to be added to the chain of \texttt{if} statements which displays the class-specific-parameters panel in case the user selects the new surface-property class in the surface-property-class-selection combo box.
In our example, the following code is suitable:
\begin{verbatim}
if(surfacePropertyString.equals(
    SURFACE_PROPERTY_DESCRIPTION_COLOURED)
  )
{
  setOptionalParameterPanelComponent(
      colourPanel
    );
}
\end{verbatim}
\end{enumerate}

For a few classes of surface property it is necessary to edit more parameters than fit into the space reserved for the surface-property panel.  
The way this has been achieved in TIM is by making the surface-property-class-specific panel consist of a button which, when clicked, initiates editing of the class-specific parameters.
The details of how this has been implemented can be seen by studying how the classes describing tiled and two-sided surfaces, \texttt{SurfaceTiling} and \texttt{TwoSidedSurface}, respectively, have been incorporated into the \texttt{SurfacePropertyPanel} class.

%\section{Adding a new type of camera}
%
%Data type \texttt{Studio} (optics.raytrace.core) holds all the information needed to render, namely information about the scene, the lights, and the camera.
%
%A ray trace is initiated by calling the \texttt{takePhoto} method of the camera.
%The camera must implement the abstract \texttt{Camera} % optics.raytrace.cameras
%class, specifically the abstract \texttt{calculatePixelColour} method, which is what the \texttt{takePhoto} method calls to calculate the colour for each pixel in the image.
%% (which also gets called by the \texttt{takePhoto} method of the \texttt{optics.raytrace.core.Studio} class).
%
%How the \texttt{calculatePixelColour} method does this depends on the camera.
%The simplest type implemented in TIM is a pinhole camera (\texttt{PinholeCamera}). % optics.raytrace.cameras
%The \texttt{calculatePixelColour} method in turn calls the scene's \texttt{getColour} method.
%This is the method that traces an individual ray backwards.

% \bibliography{/Users/johannes/Documents/work/library/Johannes}

\end{document}